\documentclass[10pt,aps,prd,twocolumn,showpacs,superscriptaddress,nofootinbib,nobibnotes,longbibliography,floatfix]{revtex4-2}

\usepackage{bm}
\usepackage{mathtools,amsmath,amssymb,amsfonts,eucal,mathrsfs,graphicx,tensor,csquotes,accents,commath,chngcntr,siunitx}
\usepackage[dvipsnames]{xcolor}
\usepackage[unicode]{hyperref}
\hypersetup{colorlinks=true, citecolor=MidnightBlue,
            linkcolor=MidnightBlue, urlcolor=MidnightBlue, linktocpage=true}
\usepackage[normalem]{ulem}

\usepackage{journals}
\usepackage{dcolumn}

\usepackage[dvipsnames]{xcolor}
\usepackage[unicode]{hyperref}
\hypersetup{colorlinks=true, citecolor=MidnightBlue,
            linkcolor=MidnightBlue, urlcolor=MidnightBlue, linktocpage=true}

\newcommand{\dhead}[1]{\multicolumn{1}{c}{$#1$}}

\usepackage{orcidlink}
\newcommand{\orcid}[1]{\href{https://orcid.org/#1}{\includegraphics[width=10pt]{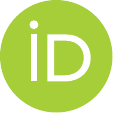}}}

\begin{document}

\title{Rapidly rotating neutron stars: Universal relations and EOS inference}

\author{Christian J. Kr\"uger \orcid{0000-0003-2672-2055}}
\email{christian.krueger@tat.uni-tuebingen.de}
\affiliation{Theoretical Astrophysics, IAAT, University of T\"ubingen, 72076 T\"ubingen, Germany}

\author{Sebastian H. V\"olkel \orcid{0000-0002-9432-7690}}
\email{sebastian.voelkel@aei.mpg.de}
\affiliation{Max Planck Institute for Gravitational Physics (Albert Einstein Institute), D-14476 Potsdam, Germany}

\date{\today}

\begin{abstract}
We provide accurate universal relations that allow to estimate the moment of inertia $I$ and the ratio of kinetic to gravitational binding energy $T/W$ of uniformly rotating neutron stars from the knowledge of mass, radius, and moment of inertia of an associated non-rotating neutron star.
Based on these, several other fluid quantities can be estimated as well.
Astrophysical neutron stars rotate to varying degrees, and, although rotational effects may be neglected in some cases, not modeling them will inevitably introduce bias when performing parameter estimation. 
This is especially important for future, high-precision measurements coming from electromagnetic and gravitational wave observations.
The proposed universal relations facilitate computationally cheap EOS inference codes that permit the inclusion of observations of rotating neutron stars.
To demonstrate this, we deploy them into a recent Bayesian framework for equation of state parameter estimation that is now valid for arbitrary, uniform rotation. 
Our inference results are robust up to around percent level precision for the generated neutron star observations, consisting of the mass, equatorial radius, rotation rate, as well as co- and counter-rotating $f$-mode frequencies, that enter the framework as data. 
\end{abstract}

\maketitle

\section{Introduction}

Neutron stars provide extreme environments that challenge our current understanding of physical theories and available methods to study them. 
They produce gravitational fields whose accurate description calls for general relativity, and the central densities require state-of-the-art nuclear and particle physics. 
These extreme conditions also allow one to constrain the unknown nuclear equation of state (EOS) at high densities and look for promising features of modified gravity theories. 
Undoubtedly, the first gravitational wave measurement of the binary neutron star merger GW170817 by the LIGO-Virgo observatories~\cite{LIGOScientific:2017vwq,LIGOScientific:2018cki} has opened a completely new window to study neutron stars in the multi-messenger era~\cite{LIGOScientific:2017ync,Sakstein:2017xjx,Ezquiaga:2017ekz,Bauswein:2017vtn,Most:2018hfd,Pacilio:2021jmq}. 
Also recently, electromagnetic observations made by NICER have allowed to put new constraints on neutron star masses and radii by modeling their hot spots~\cite{Miller:2019cac,Riley:2019yda}. 
With such ongoing successes in electromagnetic and gravitational wave astronomy that provide more precise observations in the future, accurate modeling becomes more important as well. 
This is particularly exciting for the promising capabilities of future gravitational wave detectors, such as the Einstein Telescope~\cite{Maggiore:2019uih} and Cosmic Explorer~\cite{Evans:2021gyd}; see~\cite{Huxford:2023qne,Iacovelli:2023nbv} for very recent studies.  

In order to robustly constrain the nuclear EOS from more precise observations, many aspects need to be taken into account properly. 
One main aspect is the utilization and improvement of so-called universal relations, which are a crucial tool when studying neutron star observables; they relate certain bulk quantities (or combinations thereof) of neutron stars in a way that is (largely) independent of the EOS. Among the first such relations that were discovered was a link between the normalised moment of inertia $I/MR^2$ and the compactness $C=M/R$ by Ravenhall and Pethick \cite{1994ApJ...424..846R} which was later refined by Lattimer and Prakash \cite{2001ApJ...550..426L}; similar relations were discovered and notably extended to rapidly rotating neutron stars by Breu and Rezzolla \cite{Breu:2016ufb}. Also in the 1990s, Andersson and Kokkotas discovered the first universal relations concerning mode properties, opening up the field of gravitational wave asteroseismology \cite{Andersson:1996pn,Andersson:1997rn} which has been revisited \cite{Benhar:1998au,Benhar:2004xg,Tsui:2004qd} and led to a deepened interest in such relations. Another example are the famous I-Love-Q relations that connect the moment of inertia, the Love number and the quadrupole moment in an EOS insensitive way discovered by Yagi and Yunes~\cite{Yagi:2013bca}. 
Many of the universal relations that were initially proposed have been concerned with non-rotating (or, in case the quadrupole moment $Q$ is involved, slowly rotating) neutron stars (and also quark stars) not only due to the computational complexity of constructing models of rapidly rotating neutron stars but also due to the enlarged parameter space. Latest since the late 2000s with the advance in the seismology of rotating neutron stars (cf. Gaertig and Kokkotas~\cite{2008PhRvD..78f4063G}), increasingly more universal relations have been discovered. As further examples, we mention the extension of the I-Love-Q relations to rotating stars by Pappas and Apostolatos~\cite{2014PhRvL.112l1101P} and Chakrabarti \emph{et al.}~\cite{2014PhRvL.112t1102C} or considerations regarding the general shape of rotating stars (cf. Refs.~\cite{2014ApJ...791...78A, 2021PhRvD.103f3038S, 2023ApJ...954...16G, Musolino:2023edi}). 
Traditionally, such relations have been found manually guided by intuition and experience, however, data science-driven approaches for an alternative construction of universal relations have been reported recently as well Papigkiotis and Pappas~\cite{2023PhRvD.107j3050P} and Manoharan and Kokkotas \cite{Manoharan:2023atz}.
While the majority of universal relations is concerned with isolated neutron stars in equilibrium, one should not discount the highly dynamic binary neutron star mergers and their remnants; exemplarily, we mention the universal threshold for prompt collapse discovered by Bauswein, Baumgarte, and Janka~\cite{2013PhRvL.111m1101B} which has later been revisited using simulations in full general relativity and with unequal mass binaries \cite{2021ApJ...922L..19T, 2022PhRvD.106d4026K}; there is also a link between the $f$-mode frequency of isolated NSs and the peak frequency of remnants by Lioutas, Bauswein, and Stergioulas~\cite{2021PhRvD.104d3011L}.

Universal relations not only allow one to identify and understand the relevant aspects of a problem, they are also of great importance for any related application that requires heavy computations. 
Good examples of such applications are studies that vary the underlying EOS parameters in order to compute neutron star observables for statistical parameter estimation. 
Such observables might be mass and radius, which are relatively easy to obtain for a given EOS, but also more involved properties, e.g., oscillation properties such as $f$-modes, which require a much more involved computation and cannot be easily carried out during parameter estimation. 

In this work, we provide accurate universal relations that allow to obtain properties of uniformly rotating neutron stars from the knowledge of some properties of related non-rotating neutron stars.
In a certain sense, we extend the work by Konstantinou and Morsink~\cite{2022ApJ...934..139K} who presented universal relations of rotating neutron stars for mass and radius. 
In particular, we show how the moment of inertia $I$ and the ratio of kinetic to gravitational binding energy $T/W$ at the Kepler limit can be obtained to around percent level accuracy from the  mass $M_\star$, radius $R_\star$ and moment of inertia $I_\star$ of the zero-spin star with the same central energy density $\epsilon_c$.
These values at the Kepler limit can then be used in further universal relations in order to approximate $I$ and $T/W$ for stars of arbitrary rotation rates.
Furthermore, based on the estimates for those two quantities and an estimate for the mass via the universal relation from Ref.~\cite{2022ApJ...934..139K}, we can derive estimates also for the angular momentum $J$, the rotational kinetic energy $T$, the gravitational binding energy $W$, and the proper mass $M_p$.
The advantage of the proposed universal relations is that the properties of the non-rotating star on which the universal relations are built can be easily obtained by solving one-dimensional ODEs and the universal relations themselves are simple analytic expressions; this allows computationally very cheap estimates for bulk quantities of rapidly rotating neutron stars on percent level, while the computation of the exact values (as provided, e.\,g., by the \textsc{rns} code \cite{Stergioulas:1994ea}) is computationally much more involved, as it requires the solution of an elliptic boundary value problem. 

To demonstrate the usefulness of these universal relations, we provide two applications. 
First, we apply them to recently obtained universal relations for the co- and counter-rotating $l=|m|=2$ $f$-mode frequency \cite{Kruger:2019zuz,Kruger:2020ykw}, which to date required the knowledge of the rotational neutron star properties. 
We find that the estimates for the $f$-mode frequencies are also accurate to percent level, which is required for accurate parameter estimation, see Refs.~\cite{Volkel:2021gke,Volkel:2022utc}. 
Second, we incorporate the new results into our recent Bayesian framework~\cite{Volkel:2022utc} and demonstrate the reduction of bias in inferred EOS parameters, which is important for future, high-accuracy gravitational wave measurements.  

The paper is organized as follows. 
In Sec.~\ref{sec:seq_eps_c} we discuss sequences of constant central energy density on which our discovered universal relations are based and give a brief review of their use in published literature.
In Sec.~\ref{sec:eos} we introduce our list of employed EOSs and how we calculate the sequences of neutron stars.
Section~\ref{sec:universal_relations} is devoted to the discovery of universal relations for the moment of inertia $I$ and the ratio of kinetic to gravitational binding energy $T/W$ for rapidly (and uniformly) rotating neutron stars.
In Sec.~\ref{sec:application} we explain how the co- and counter-rotating $l=|m|=2$ $f$-mode frequencies of rapidly rotating neutron stars may be estimated when combined with previously published universal relations \cite{Kruger:2019zuz} and we discuss the accuracy of these estimates.
As a second application, we extend our previous EOS inference code \cite{Volkel:2021gke} to allow for observations of rapidly rotating neutron stars by implementing the required universal relations.
We consider three test cases and discuss in which of these it is crucial to account for rotational corrections. 
Last, we summarise our findings in Sec.~\ref{sec:conclusions} and discuss limitations and potential extensions. 
Unless otherwise noted, we employ units in which $c = G = M_\odot = 1$ throughout this paper.

\section{Sequences of constant central energy density}
\label{sec:seq_eps_c}

In this paper, we focus on sequences of neutron stars along which the central energy density $\epsilon_c$ is held constant and the angular rotation rate $\Omega = 2\pi f_{\rm spin}$ is varied. For a given $\epsilon_c$, such a sequence is bounded by the non-rotating limit ($\Omega = 0$) and the mass-shedding (or Kepler) limit where the neutron star rotates at the maximum possible angular rotation rate $\Omega = \Omega_K$. Here and henceforth, we use the subscript ``$K$'' to denote a quantity that belongs to a model rotating at the Kepler limit; further, we use the subscript ``$\star$'' for quantities of the non-rotating member of the sequence.

While such sequences may have only limited astrophysical motivation (during the spin-down of an isolated neutron star, its baryon mass will remain constant but not its central energy density), they are numerically extremely simple to generate. After specifying an EOS, the widely used \textsc{rns} code \cite{Stergioulas:1994ea}, which we employ to construct rotating equilibrium configurations, takes a central energy density $\epsilon_c$ and some axis ratio $\mathfrak{r} = r_p / r_e$ (where $r_p$ and $r_e$ are the polar and equatorial coordinate radii, respectively) as parameters and then finds the corresponding solution to the Einstein field equations; constructing sequences of, e.\,g., constant baryon mass is computationally more expensive and requires an iterative search for the corresponding pair $(\epsilon_c, \mathfrak{r})$. The conceptual simplicity of sequences of constant central energy density is one reason why these have been used in several studies (e.\,g.,~Refs.~\cite{2004MNRAS.352.1089S, 2006MNRAS.368.1609D}). Furthermore, several universal relations that are based on such sequences have been discovered; these are relations for the $f$-mode frequencies and damping times \cite{2008PhRvD..78f4063G, 2011PhRvD..83f4031G, Kruger:2019zuz} and those for mass and radius \cite{2022ApJ...934..139K}.

The latter study has revealed a curious insensitivity to EOSs in rapidly rotating neutron stars: The knowledge of only mass $M_\star$ and radius $R_\star$ of the non-rotating member of the sequence allows to estimate to high accuracy the mass and equatorial radius of a star rotating at a given angular rotation rate $\Omega$ using simple analytical formulae. As has been pointed out in Ref.~\cite{2022ApJ...934..139K}, such universal relations may be used in EOS inference codes when working with observations of (sufficiently) rapidly rotating neutron stars; the requirement to construct rotating equilibrium configurations by means of, e.\,g., the \textsc{rns} code can be circumvented by solving the much simpler TOV equations and then applying the universal relations. However, as of yet, this method is limited to mass as well as radius and also the $f$-mode frequency as a perturbative quantity.

In a recent work \cite{Volkel:2021gke}, we approached the EOS inference problem for observations of slowly rotating stars for which the departure from sphericity may be neglected (hence mass and radius can safely be assumed to be those acquired by the TOV solver without correction). In order to include a potentially observed $f$-mode frequency in the EOS inference, we used a universal relation to account for rotational corrections in these; however, we still (wrongly) used the properties of a non-rotating star in this universal relation as a good approximation. 

We extend our previous work to faster spinning neutron stars by building on sequences of constant central energy density; they turn out to provide a useful slicing of the $M$-$R$-plane (given a specific EOS, we refer to the area in an $M$ vs. $R$ diagram that is covered by dynamically stable, rotating equilibrium configurations of neutron stars as the $M$-$R$-plane) for which EOS insensitive relations for bulk quantities of neutron stars can be found. We would not expect to find similarly simple relations if we, e.\,g., considered sequences of constant gravitational or baryon mass, since for these sequences one parameter (namely the mass) is held constant and, hence, cannot provide information with respect to the sequence. One might imagine yet different sequences of neutron stars (or, equivalently, a different slicing of the $M$-$R$-plane), but we opt to focus on sequences of constant central energy density owing to their conceptual simplicity.

\section{Equations of State}
\label{sec:eos}

We employ piecewise polytropic (PP) equations of state \cite{Read:2008iy} in our study as the four parameters $(p_1, \Gamma_1, \Gamma_2, \Gamma_3)$ used in this approximation allow us to easily cover a wide range of EOSs.
We note that the choice of this parametrisation may introduce a bias as the set of PP EOSs is not dense within the set of all physically allowed EOSs; however, it has been observed that universal relations which are based on the PP parametrisation are robust and the introduced bias is very small \cite{2022ApJ...934..139K}.

In order to reveal relations for bulk quantities of neutron stars that are (mostly) insensitive to the EOS choice, we pick 31 arbitrary PP EOSs of those that have been fitted to tabulated EOSs (cf.~Refs.~\cite{Read:2008iy, 2019PhRvD..99l3026K} for tables of coefficients).
We currently do not consider speed of sound constraints \cite{2018ApJ...860..149T, 2022ApJ...939L..34A}, however, the universality of the relations is still remarkable and should improve with a narrower set of EOSs; we also note that our selection of EOSs does not account for phase transitions, which potentially have interesting features in binary neutron star mergers (e.g., Ref.~\cite{Weih:2019xvw}).

\begin{figure}
    \centering
    \includegraphics[width=1.0\linewidth]{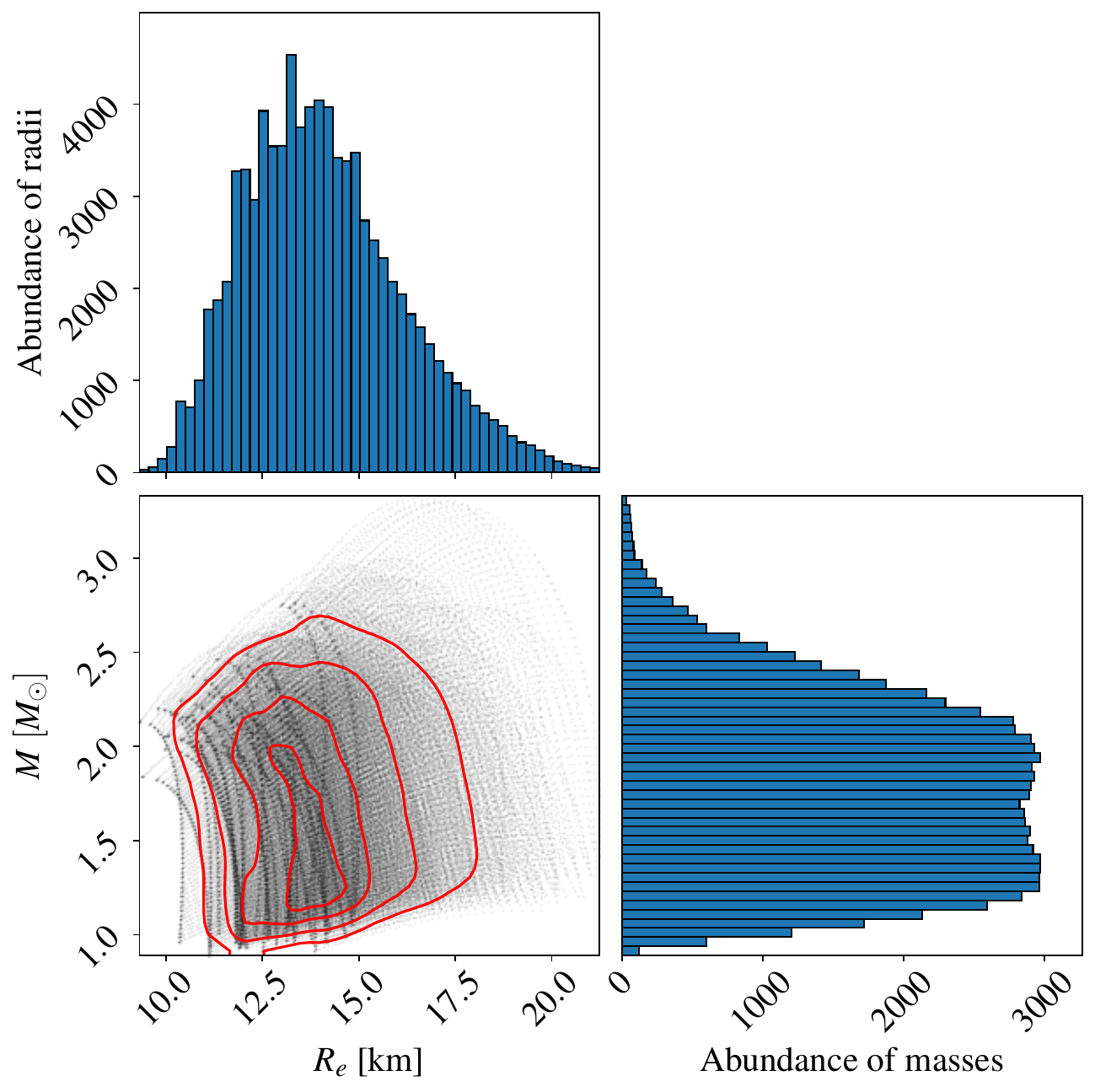}
    \caption{A corner plot of the dataset on which we construct the universal relations. Each black dot in the $M$ vs. $R$ diagram represents one of 83011 NS models in our dataset. Their density distribution is indicated by the red contour lines. The top and right diagram show the corresponding histograms of radius and mass, respectively. The darker lines correspond to the $M$-$R$-curves of the EOSs which are visible because our dataset contains some additional near-spherical models; see the main text for more details.}
    \label{fig:mr_hist2d}
\end{figure}

Using these EOSs, we construct 1550 sequences of constant central energy density that comprise 83011 individual neutron star models. 
We pick the lowest $\epsilon_c$ for an EOS as an integer multiple of $0.05 \times 10^{15}\,\unit{g}\,\unit{cm}^{-3}$ such that the maximally rotating (and at the same time the heaviest) member of that sequence has a mass of at least $1.1\,M_\odot$; this way, we ensure that our universal relations are calibrated for neutron stars that are even a bit lighter than astrophysically expected \cite{2018MNRAS.481.3305S, 2015ApJ...812..143M}.
For each EOS, we construct 50 sequences; we choose the corresponding central energy densities such that the non-rotating stars of one particular EOS are roughly evenly spaced in their masses. The models along a sequence are evenly spaced in the axis ratio $\mathfrak{r}$ with $\Delta \mathfrak{r} = 0.01$; in addition, we constructed several near-spherical models with axis ratios $\mathfrak{r} \in [0.99, 1.00]$ (which corresponds to spin frequencies up to $\sim 200-300\,$Hz) in order to have a good resolution also for slowly rotating models. Last, any sequence also contains the Keplerian model, which, in general, does not fit into the evenly spaced grid. Hence, every sequence contains about 55 neutron star models, ranging from zero to maximal rotation. We visualise our dataset in Fig.~\ref{fig:mr_hist2d} using a corner plot for mass and radius.

The maximum masses of spherical models resulting from these EOSs span the range of $1.83\,M_\odot$ to $2.75\,M_\odot$; the radii of non-rotating stars with a gravitational mass of $1.4\,M_\odot$ range from $10.4\,$km to $14.9\,$km. These ranges demonstrate that we include EOSs that are well outside the current astrophysical constraints (see, e.\,g., Refs.~\cite{2013Sci...340..448A, 2021ApJ...908..122G, 2020Sci...370.1450D, 2021ApJ...918L..29R}) but we include them nonetheless in order to prove the robustness of our universal relations on an even broader range of EOSs.

\section{Universal Relations}
\label{sec:universal_relations}

\subsection{Moment of Inertia}

Along a given sequence of neutron stars from zero to maximal rotation, the moment of inertia $I$ varies by definition from $I_\star$ to $I_K$. As both mass and (equatorial) radius of the stars increase along a sequence, it is intuitively clear that also the moment of inertia monotonically increases. We map the interval $[ I_\star, I_K ]$ onto the unit interval $[0, 1]$ by considering the linearly rescaled moment of inertia $I_n := (I - I_\star) / (I_K - I_\star)$. In the same manner, we normalise a star's rotation rate $\Omega$ by its corresponding Kepler limit $\Omega_K$ and introduce the fractional rotation rate $\Omega_n := \Omega / \Omega_K$ which then also spans the unit interval.

\begin{figure}
    \centering
    \includegraphics[width=1\linewidth]{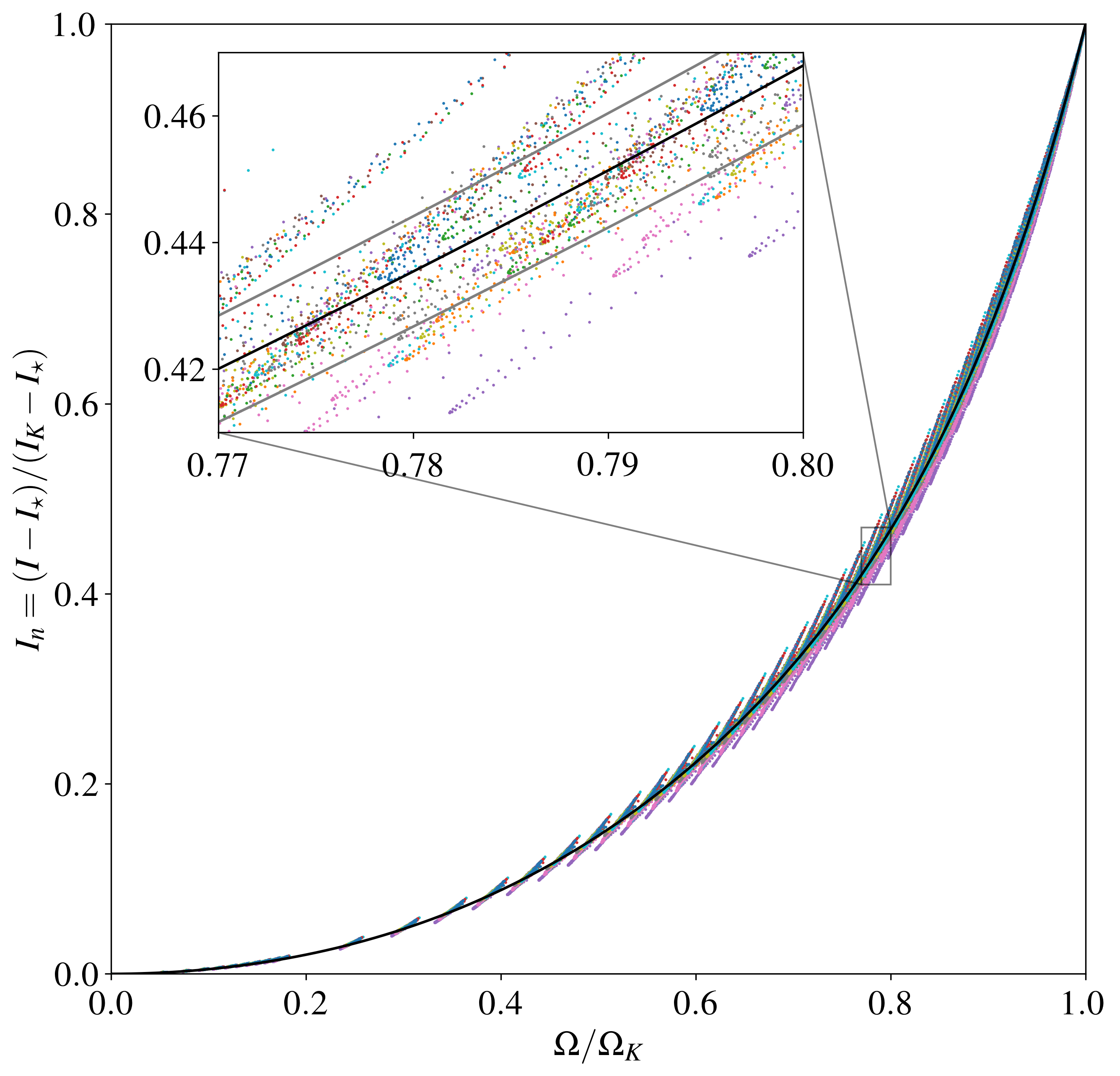}
    \caption{A scatter plot of the rescaled moment of inertia $I_n$ against the fractional angular rotation rate $\Omega_n$ for the 1550 sequences of stars in our dataset. As the graph shows a very large number of data points, we display an inset, in which we enlarge a small region of the graph; it becomes visible that the majority of the data points gather within 2\% (indicated by grey lines, not shown in the main graph) of the polynomial fit (black solid line) and there are some with a larger deviation from the fit. Upon inspection it turns out that the latter belong to neutron stars with rather small masses $M \lesssim 1.2\,M_\odot$. The relative error may appear large in the inset; however, the fit is made for the rescaled quantity $I_n$ which suffers from truncation errors. The relative error for the moment of inertia $I$ is bounded by $1.9\%$; see the discussion in the main text for details.}
    \label{fig:I_vs_Om}
\end{figure}

Having rescaled the moments of inertia and the stars' angular rotation rates of our dataset, we show a scatter plot of these in the unit square in Fig.~\ref{fig:I_vs_Om}. The angular rotation rate $\Omega$ and the moment of inertia $I$ are normalised in such a way that the curves of any sequence will pass by construction through the points $(0, 0)$ and $(1, 1)$, which correspond to the non-rotating star and the star at the Kepler limit, respectively. Such a rescaling to the unit interval has proven to be very useful in related works, e.g., in Ref.~\cite{2022ApJ...934..139K} or for studying the sound speed in Ref.~\cite{Ecker:2022xxj}. However, we find that not only these two points belong to the graph of a sequence but, in fact, the entire graph connecting those two points is (with only minor deviations) independent of the EOS as well as the central energy density that is chosen for a particular sequence. The figure strongly suggests a relation $I_n \approx f(\Omega_n)$ and we propose a simple polynomial fit of the form
\begin{align}
    f(\Omega_n)
    & = \sum_{k=1}^4 c_{2k} \Omega_n^{2k}.
    \label{eq:ur_Omega_n}
\end{align}
We omit the constant term in order to enforce $f(0) = 0$ and we account only for terms of even power since the corrections to the moment of inertia $I_\star$ of the non-rotating star are of quadratic order. When fitting the coefficients to the data, we also tried odd powers of $\Omega_n$ in the fitting function but the odd coefficients either were afflicted with large error bars or resulted in larger deviation from the data points at low rotation rates. The coefficients we propose to use for the fit are shown in Table~\ref{tab:coeff_unit_square} (the coefficients resulting from a least-squares fit sum up to 0.9986; we rescale these linearly so that their sum is 1.0). The moment of inertia $I$ of a star rotating at the (fractional) angular rotation rate $\Omega_n$ can then be estimated by the formula
\begin{align}
    I(\Omega_n)
    & = I_\star + \left(I_K-I_\star \right) f\left(\Omega_n \right).
    \label{eq:ur_I}
\end{align}

\begin{table*}
    \caption{Coefficients used in Eq.~\eqref{eq:ur_Omega_n} which is used in the universal relations for $I$ and $T/W$.}
    \label{tab:coeff_unit_square}
    \begin{ruledtabular}
    \begin{tabular}[v]{lcddddcc}
        Equations & Quantity & \dhead{c_2} & \dhead{c_4} & \dhead{c_6} & \dhead{c_8} & Max err. & Mean err. \\
        \hline
        \eqref{eq:ur_Omega_n}, \eqref{eq:ur_I} & $I$   & 0.4864 & 0.4542 & -0.4218 & 0.4797 & 1.8\% & 0.24\% \\
        \eqref{eq:ur_Omega_n}, \eqref{eq:ur_TW} & $T/W$ & 0.7842 & 0.1905 & -0.1056 & 0.1305 & 4.3\% & 0.57\%   
    \end{tabular}
    \end{ruledtabular}
\end{table*}

We show the polynomial fit from Eq.~\eqref{eq:ur_Omega_n} in Fig.~\ref{fig:I_vs_Om}, too. In the inset, we enlarge on a very small region of the graph in order to visualise the structure of the data points; it is apparent that the majority of the data points lie very close the fitting curve and only a small portion of data points show a larger deviation from the fit. Upon inspecting those data points, we find that the latter belong to neutron stars with rather small masses $M \approx 0.9 - 1.2\,M_\odot$. We observe that this range of masses is roughly the threshold below which universality is marginally reduced and the sequences tend to deviate increasingly more from the main band visible in the inset. In general, the fit is quite tight and independent of the EOS and central energy density. The fit is based on 83011 NS models constructed using 31 different piecewise polytropes.

It is important to note that the relative error\footnote{In Sec.~\ref{sec:universal_relations} for the newly proposed universal relations for $I$ and $T/W$, we define the relative error to be $|1 - y/y_{\rm{fit}}|$, i.e., relative to the fit.} in Fig.~\ref{fig:I_vs_Om} may appear very large; indeed, in the inset, the relative error is up to 7\% and for lower $\Omega_n$ it can grow naturally (due to small absolute values) to more than 50\%. This is an artefact of the finite numerical accuracy when constructing equilibrium models (we demand an accuracy of $10^{-6}$ in the \textsc{rns} code); the impact grows largest for slowly rotating models where the difference $I - I_\star$ suffers from truncation errors. However, the fit is made for the rescaled quantity $I_n$. When we compare the correct $I$ to its estimate via Eq.~\eqref{eq:ur_I}, then the largest relative error in our dataset is $1.8\%$ and the mean relative error of the estimates is $0.24\%$.

\begin{figure}
    \centering
    \includegraphics[width=1\linewidth]{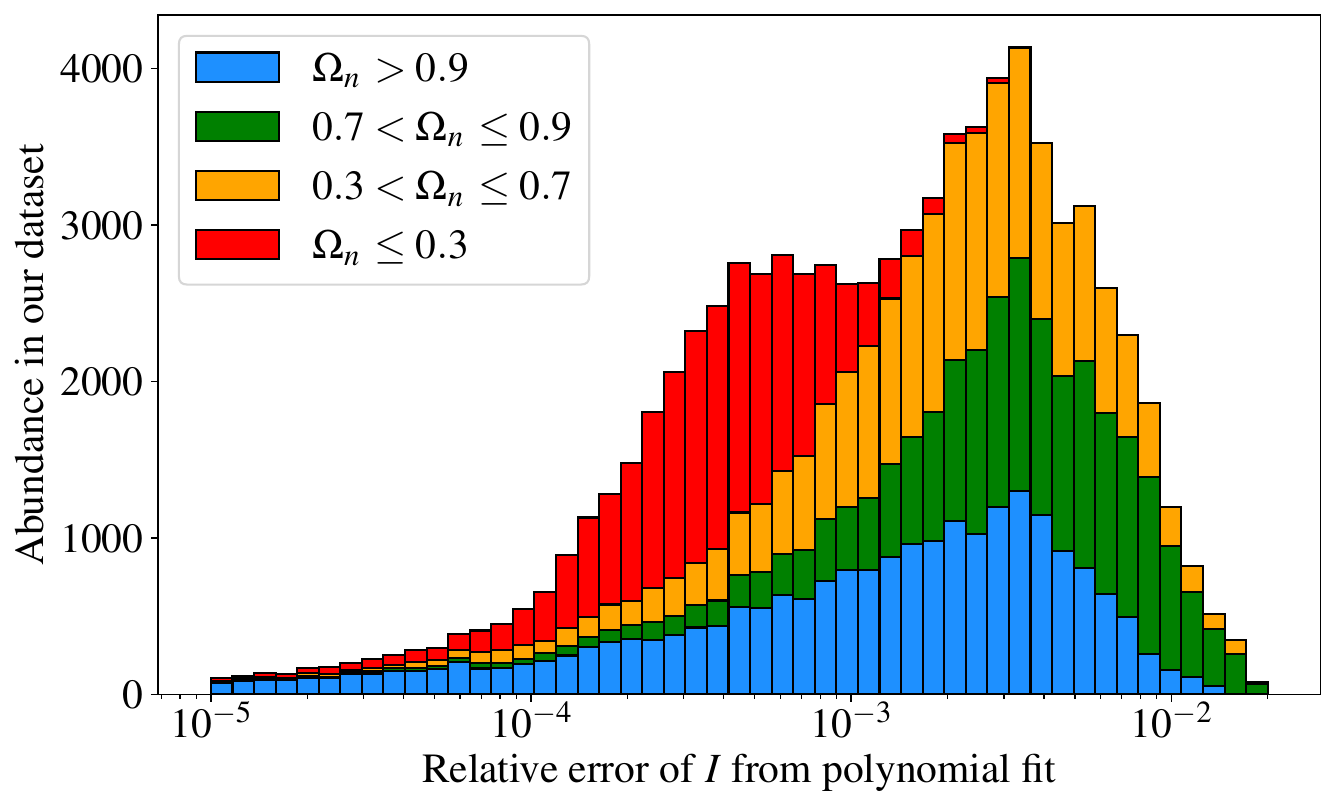}
    \caption{A stacked histogram of the relative deviations of our 83011 data points from the universal relation Eq.~\eqref{eq:ur_I}. The majority of data points (roughly $80\%$) deviate less than $0.4\%$ from the fit and the maximum relative error is $1.9\%$. The relative errors are subdivided corresponding to the fractional rotation rate $\Omega_n$ of the neutron star model; this shows that the largest deviations from the correct value appear for rotation rates $\Omega_n \approx 0.7 - 0.9$.}
    \label{fig:I_vs_Om_hist}
\end{figure}

Since we have well above 80000 data points, we do not show the relative error as a function of the rotation rate in a cluttered scatter plot but instead show a histogram of the relative errors in Fig.~\ref{fig:I_vs_Om_hist}. We note that $80\%$ of the models in our dataset deviate less than $0.4\%$ from the estimate, and $97.3\%$ are fitted better than $1\%$. The largest errors occur for rather large rotation rates $\Omega_n \approx 0.7 - 0.9$ (this can be seen in Fig.~\ref{fig:I_vs_Om} where the bulge of the data points is largest and in the stacked histogram in Fig.~\ref{fig:I_vs_Om_hist} where the corresponding rotation rates are shown in green). Intuitively, higher rotation rates will favour some deviation from universality; however, as the sequence will also have to pass exactly through the point $(1, 1)$, we expect that the sequences will naturally show more universal behaviour close to this point and hence the fit becomes more precise close to the Kepler limit.

\begin{table*}
    \caption{Coefficients $d_k$ and the combined quantities $x$ and $y$ that are used in Eq.~\eqref{eq:universal_x_y}.}
    \label{tab:coeff_kepler}
    \begin{ruledtabular}
    \begin{tabular}[v]{ccdddddcc}
        $x$ & $y$ & \dhead{d_0} & \dhead{d_1} & \dhead{d_2} & \dhead{d_3} & \dhead{d_4} & Max err. & Mean err. \\
        \hline
        $I_\star R_\star^{-3}$ & $I_K R_\star^{-3} C_\star^3$ & -2.661 & 0.0   & -0.6221 & 0.03786 & 0.01445 & 2.2\% & 0.55\% \\
        $I_\star M_\star^{-3}$ & $\left(T/W\right)_K C_\star^{-4}$ & -6.338 & 9.808 & -3.853  & 0.8878  & -0.07858  & 3.2\% & 0.59\%
    \end{tabular}
    \end{ruledtabular}
\end{table*}

\subsection{Moment of Inertia at the Kepler limit}

The universal relation proposed in Eq.~\eqref{eq:ur_I} allows to estimate the moment of inertia of a rotating neutron star, provided that the two moments of inertia $I_\star$ and $I_K$ at both ends of the sequence are known. While $I_\star$ is relatively easily accessible via Hartle's slow-rotation formalism \cite{1967ApJ...150.1005H}, the value of $I_K$ requires a comparatively large computational expense for its precise calculation, e.\,g., the iterative solution for the Kepler model employing the \textsc{rns} code. However, instead of iterating for the Kepler limit, one might as well iterate for the desired rotation rate and directly find the sought-after moment of inertia.

Instead of employing an external code to solve the Einstein field equations in order to find $I_K$, we will propose a universal relation that provides a very good estimate to $I_K$ based on mass, radius and moment of inertia of the non-rotating star (keeping in mind that we always work with sequences of constant central energy density). For this universal relation, we define the two auxiliary quantities
\begin{equation}
    x := \frac{I_\star}{R_\star^3}
    \qquad\text{and}\qquad
    y := I_K \frac{C_\star^3}{R_\star^3},
    \label{eq:def_xy_I}
\end{equation}
where $C_\star := M_\star / R_\star$ is the compactness of the non-rotating star.

\begin{figure}
    \centering
    \includegraphics[width=1.0\linewidth]{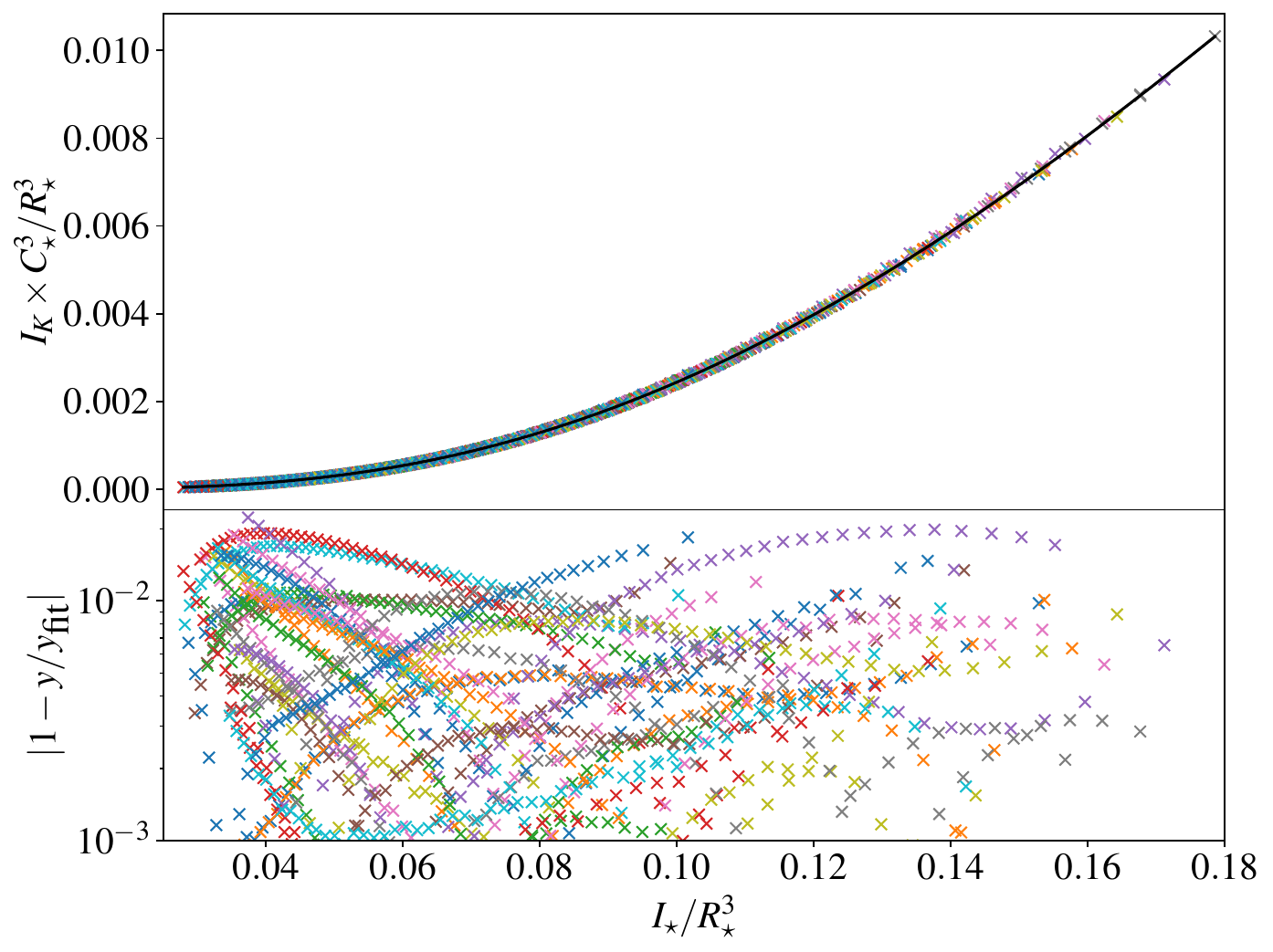}
    \caption{The scaled moment of inertia of neutron stars at the Kepler limit against the radius-scaled moment of inertia of the non-rotating neutron star with the same central energy density as the rotating stars. The bottom panel shows the relative deviation of the data points from the fit (black solid line); the maximum relative error is $2.2\%$.}
    \label{fig:logIK_of_logstar}
\end{figure}

In Fig.~\ref{fig:logIK_of_logstar}, we show a scatter plot of $y$ vs. $x$ for the 1550 sequences of constant central energy density that we computed; the points lie nearly perfectly on a line which we fit using a functional expansion of the type
\begin{align}
    \ln y
    & = \sum_{k = 0}^4 d_k \left( \ln x \right)^k.
    \label{eq:universal_x_y}
\end{align}
We show the coefficients of a least-squares best-fit in Table~\ref{tab:coeff_kepler} (we have annulled $d_1$ manually as it would be afflicted with a large uncertainty). The accuracy of the fit is visualised in the lower part of the same graph; the maximum relative deviation from the universal relation in our dataset is $2.2\%$.

Using this relation, it becomes clear that we can get a very good estimate for $I_K$ out of the triple $(M_\star, R_\star, I_\star)$; for a given $\epsilon_c$, these three values can relatively easily be computed using the TOV equations along with Hartle's equation. Combined with Eq.~\eqref{eq:ur_I}, we can estimate the moment of inertia of a neutron star spinning at a fractional rotation rate $\Omega_n$ to high accuracy. In a more realistic setting, one would be interested in the moment of inertia of a star rotating at a specific angular rotation rate $\Omega$ rather than at a certain fractional rotation rate $\Omega_n$. This is no restriction since the Kepler velocity $\Omega_K$ that links those two can also be estimated well from $M_\star$ and $R_\star$ via Eq.~(1) from Ref.~\cite{2022ApJ...934..139K}.

\subsubsection{Normalisations for the moment of inertia}

We comment briefly on our auxiliary quantities $x$ and $y$; apparently, we construct quantities of the form $\sim I / R^3$ and discover universal relations between these. This is different from several studies in which the moment of inertia is considered in other combinations. Probably the most intuitive normalisation is based on the expressions for the moment of inertia of axisymmetric objects as found in Newtonian theory; there, the moment of inertia is proportional to ``mass times square of radius'' and, hence, several authors (e.g., Refs.~\cite{1994ApJ...424..846R, 2005ApJ...629..979L}) linked the \emph{dimensionless moment of inertia} $\tilde{I}:= I / (MR^2)$ of neutron stars to its compactness $C = M/R$. Later, it was discovered that employing the \emph{mass-normalised moment of inertia} $\bar{I} := I / M^3$ allows for tighter universal relations regarding the $f$-mode frequency \cite{2010ApJ...714.1234L} and, subsequently, the universal relations for the moment of inertia as a function of the compactness were updated \cite{Breu:2016ufb}. Indeed, we can report that the auxiliary quantities $x' = \bar{I} = I_\star / M_\star^3$ and $y' = I_K / (M_\star^3 C_\star^2)$, in both of which the mass-normalised moment of inertia can be spotted, also obey a universal relation; however, with a relative error of up to $4\%$, it is not as tight as the universal relation between the auxiliary quantities $x$ and $y$ when they are loosely based on $I / R^3$ as we defined them in Eq.~\eqref{eq:def_xy_I}. For a more thorough discussion of the history of the combinations in which the moment of inertia appears in universal relations and their physical motivation, see Sec. 4.1 in Ref.~\cite{Breu:2016ufb}.

\subsection{\texorpdfstring{Ratio of kinetic to gravitational binding energy $T/W$}{bla}}

\begin{figure}
    \centering
    \includegraphics[width=1\linewidth]{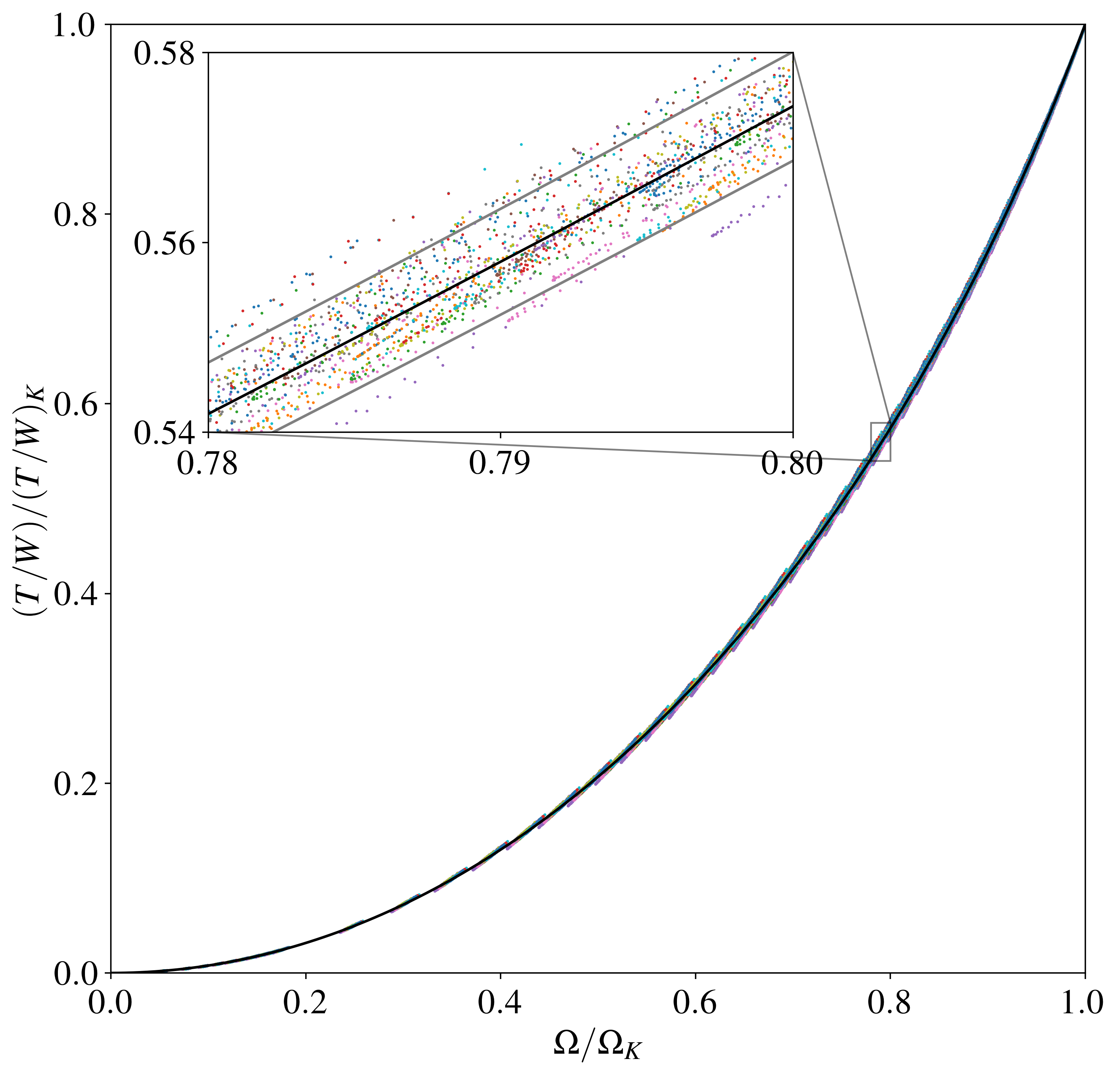}
    \caption{A scatter plot of the rescaled $(T/W)_n$ against the fractional angular rotation rate $\Omega_n$ for the 1550 sequences in our dataset. An inset shows an enlargement of a small region of the graph; almost all data points are inside a very narrow band (with less than $1\%$ relative error which is indicated by grey lines; not shown in the main graph) around the polynomial fit (black solid line). The few individually visible points outside the dense band have a relative error of less than $2\%$.}
    \label{fig:TW_vs_Om}
\end{figure}

\begin{figure}
    \centering
    \includegraphics[width=1.0\linewidth]{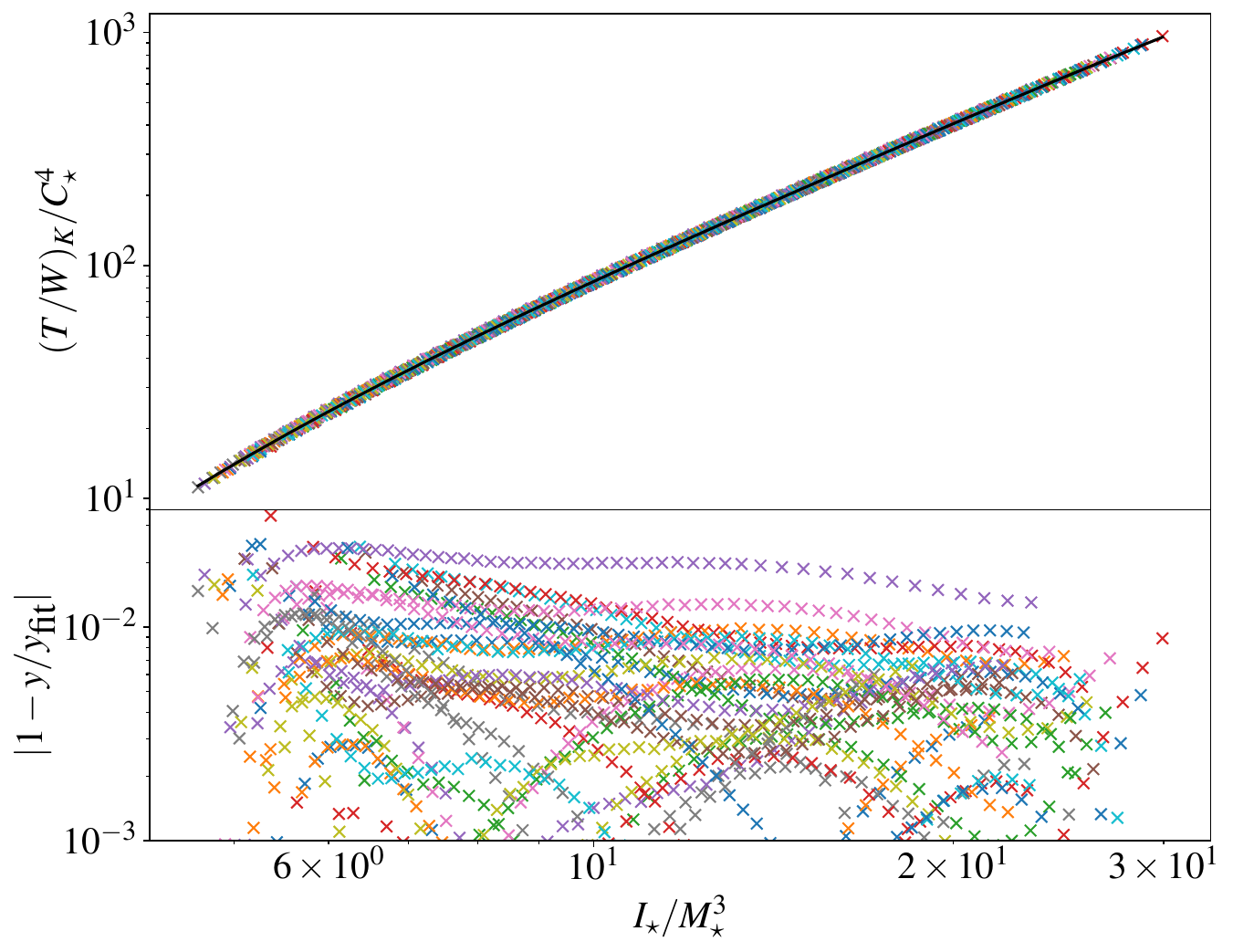}
    \caption{The scaled value of $T/W$ of neutron stars at the Kepler limit against the mass-scaled moment of inertia of the non-rotating neutron star with the same central energy density as the rotating stars. The bottom panel shows the relative errors of the data points; the maximum relative deviation from the best fit is $3.2\%$.}
    \label{fig:logTWK_of_logstar}
\end{figure}

In the previous sections, we have demonstrated how the moment of inertia of a rotating star can be estimated out of the bulk quantities of the non-rotating star (with the same central energy density). We find that in structurally the same way, it is also possible to estimate the value of $T/W$, the ratio of rotational kinetic to gravitational binding energy\footnote{Even though a binding energy (in our case the gravitational binding energy $W$) is usually considered negative, since this is the energy that has to be \emph{added} to a system in order to put infinite separation between its individual particles, we follow the definition of Refs.~\cite{1986ApJ...304..115F, 1992ApJ...398..203C} whereby we have $W > 0$. This allows us to omit the absolute value bars in $T/|W|$ and declutter the formulae in the text; we only need to keep in mind that $W$ denotes the negative of the (negative) binding energy.}, of a rotating star. Even though $T/W$ might be difficult to observe, it is worthwhile having a universal relation for it since we can use it to derive several other bulk quantities as we will argue at the end of this section. In order to find the universal relation, we normalise $T/W$ in the same way as we normalised the moment of inertia $I$; keeping in mind that $(T/W)_\star$ vanishes since $T \propto \Omega^ 2$, we get a slightly simpler expression $(T/W)_n := (T/W) / (T/W)_K$. Again, a scatter plot shown in Fig.~\ref{fig:TW_vs_Om} suggests a very tight relation $(T/W)_n \approx f(\Omega_n)$ and we use the same polynomial expression given in Eq.~\eqref{eq:ur_Omega_n} as fitting function. The best-fit coefficients are shown in Table~\ref{tab:coeff_unit_square} (after rescaling their sum from 0.9996 to 1.0). A very good approximation for the value of $T/W$ of a star rotating at the fractional rotation rate $\Omega_n$ is then given by

\begin{align}
    (T/W)(\Omega_n)
    & = \left(T/W\right)_K f\left( \Omega_n \right).
    \label{eq:ur_TW}
\end{align}
As before, in order to make use of this formula, we need to know $(T/W)_K$, i.e., the value of $T/W$ at the Kepler limit (in our dataset this value ranges in the interval $[0.084, 0.146]$, i.e., considerably below the value of $\sim 0.255$ for the onset of the bar-mode instability that requires differential rotation \cite{2007PhRvD..75d4023B}). We find that a very tight relation between the two auxiliary quantities
\begin{equation}
    x := \frac{I_\star}{M_\star^3}
    \qquad\text{and}\qquad
    y := \left(T/W\right)_K C_\star^{-4}
    \label{eq:def_xy_TW}
\end{equation}
exists. The relation between those two can be well modeled by the functional dependence given in Eq.~\eqref{eq:universal_x_y} and the corresponding least-squares best-fit coefficients are listed in Table~\ref{tab:coeff_kepler}. We show a scatter plot of our data points in Fig.~\ref{fig:logTWK_of_logstar}.

\begin{figure}
    \centering
    \includegraphics[width=1\linewidth]{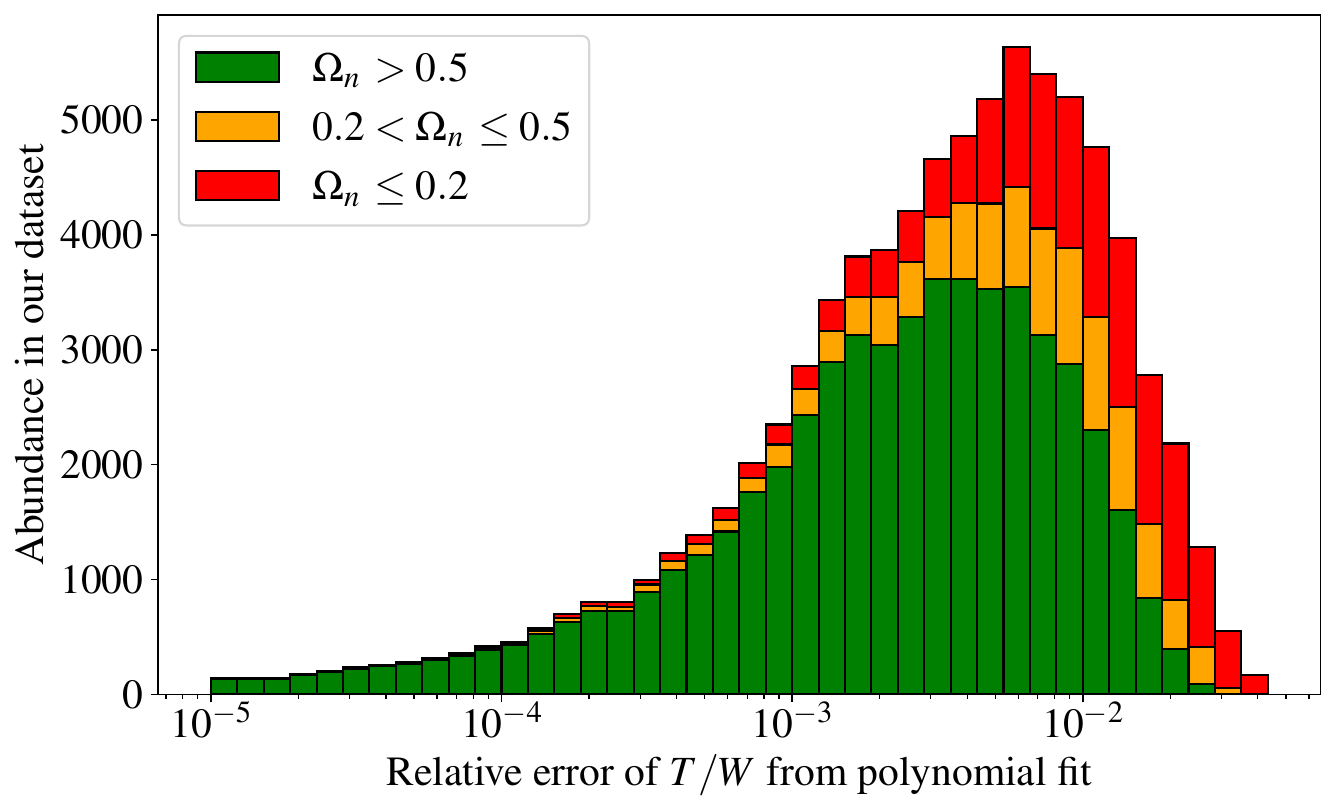}
    \caption{A stacked histogram of the relative errors of the universal relation for $T/W$; they remain below $4.3\%$ for all models in our dataset. The relative errors are subdivided corresponding to the fractional rotation rate $\Omega_n$ of the neutron star model; this shows that the deviation from the correct value is largest for models with a small rotation rate since the absolute value of $T/W$ is small for these.}
    \label{fig:TW_rel_err}
\end{figure}

The relative error of the proposed universal relation for $T/W$ is less than $4.3\%$ for the neutron star models in our dataset. It is worth noting that for slowly rotating models the absolute value of $T/W$ is small and hence the relative error of the estimate is the largest. We show a stacked histogram of the relative errors in Fig.~\ref{fig:TW_rel_err}, where we subdivide the relative errors corresponding to the fractional rotation rate; it is apparent that the universal relation becomes more accurate for larger rotation rates.

We conclude the description of the universal relations by noting that at the same time we get accurate estimates for the angular momentum $J$, the rotational kinetic energy $T$, and the gravitational binding energy $W$ as well as the proper mass $M_p$ of rapidly rotating neutron stars as freebies. Having estimated $M$ (as described in Ref.~\cite{2022ApJ...934..139K}), $I$ and $T/W$ via universal relations, we can then calculate $J = I \Omega$, $T = \frac{1}{2} I \Omega^2$, $W = T / (T/W)$ and finally $M_p = W + M - T$. We have not conducted a thorough analysis of the accuracy yet, but initial tests show that the relative errors of those four quantities are comparable to those of $I$ and $T/W$.

\section{Application and results}
\label{sec:application}

In this section we demonstrate the capabilities of the new universal relations when combined with those proposed in Ref.~\cite{2022ApJ...934..139K} for the mass and equatorial radius. 
In Secs.~\ref{sec:fmode_ctr} and \ref{sec:fmode_cor} we show the accuracy when applied to the universal relations for the co- and counter-rotating $l=|m|=2$ $f$-mode frequencies from Refs.~\cite{Kruger:2019zuz,Kruger:2020ykw}. 
In Sec.~\ref{sec:bayes_inf} we incorporate them into the Bayesian EOS framework from Ref.~\cite{Volkel:2022utc}, which demonstrates its capacities to perform quick inference for generic neutron star observables beyond slow-rotation approximations.

\subsection{\texorpdfstring{Counter-rotating $f$-mode frequency}{bla}}
\label{sec:fmode_ctr}

In addition to mass, radius, and moment of inertia, we are also interested in using the $f$-mode frequencies as potential observables in the Bayesian framework. In order to do so, we need to estimate these also from quantities of the non-rotating star. The work presented in Ref.~\cite{Kruger:2019zuz} offers (when combined with the above proposed universal relation for the moment of inertia, cf.~Eq.~\eqref{eq:ur_I}) two ways of approximating the $f$-mode frequencies of a star with a given central energy density and rotation rate; as both approximations differ slightly in accuracy, we will explain and employ both of them.

As we rely heavily on sequences of constant central energy density, the most obvious way to find the $f$-mode frequencies is to employ Eq.~(4) of Ref.~\cite{Kruger:2019zuz} which is also fitted to the same type of neutron star sequences. We reproduce the fitting function for completeness:\footnote{In the cited work (Ref.~\cite{Kruger:2019zuz}), an index ``i'' is used to indicate that the $f$-mode frequency is measured in an inertial frame as opposed to a corotating frame. In the present work, we work solely with frequencies in the inertial frame and hence we omit the index ``i'' here.}
\begin{equation}
    \frac{\sigma^\textrm{x}}{\sigma_\star}
    = 1 + a_1^\textrm{x} \left( \frac{\Omega}{\sigma_\star} \right)
        + a_2^\textrm{x} \left( \frac{\Omega}{\sigma_\star} \right)^2,
    \label{eq:eq4_kk20}
\end{equation}
where the index $\textrm{x}$ is used to distinguish between the stable co-rotating branch (labelled $\textrm{s}$) and potentially unstable counter-rotating branch (labelled $\textrm{u}$) of the $f$-mode; the coefficients $a_1^\textrm{x}$ and $a_2^\textrm{x}$ are reported in the same work. Given the $f$-mode frequency $\sigma_\star$ of the non-rotating star of the sequence, this universal relation provides an estimate for the $f$-mode frequencies of a star rotating with the angular rotation rate $\Omega$. 

Effectively, we have reduced the problem to determining the $f$-mode frequency $\sigma_\star$ of the non-rotating star which is a comparatively easy task. We might want to use $\sigma_\star$ from our dataset based on time evolutions which was built in preparation for the work that led to the discovery of $f$-mode universal relations \cite{Kruger:2019zuz,Kruger:2020ykw}. This would be consistent as these universal relations are also calibrated using those values. However, since the EOS database for the previous work has focused on rotating neutron stars, it does by far not contain enough values for $\sigma_\star$; we need this value essentially for any non-rotating neutron star of an arbitrary EOS. Instead, we will use a very accurate universal relation discovered by Lau, Leung, and Lin~\cite[Eq. (6)]{2010ApJ...714.1234L}\footnote{Note that Lau, Leung, and Lin~\cite{2010ApJ...714.1234L} denote the $f$-mode frequency by $\omega_r$ which is related to our $f$-mode frequency $\sigma_\star$ via $\omega_r = 2\pi\sigma_\star$.} to estimate $\sigma_\star$ from the mass $M_\star$ and the moment of inertia $I_\star$; our database (as described in Sec.~\ref{sec:eos_db}) contains those values. The most accurate value for $\sigma_\star$ could be reliably found by using the estimate for $\sigma_\star$ from the universal relation as an initial guess for the (numerically) exact solution via an eigenvalue code; however, this would considerably increase the computational expense of the problem in return for an unnecessarily accurate value of the $f$-mode frequency: Keep in mind that the universal relation for the $f$-mode frequency of rotating stars works only to percent level; furthermore, we also assume that the observed $f$-mode frequencies are tainted with error bars of several percent. The subpercent accuracy of the universal relation proposed in Ref. \cite{2010ApJ...714.1234L} is certainly sufficient.

\begin{figure}
    \includegraphics[width=1.0\linewidth]{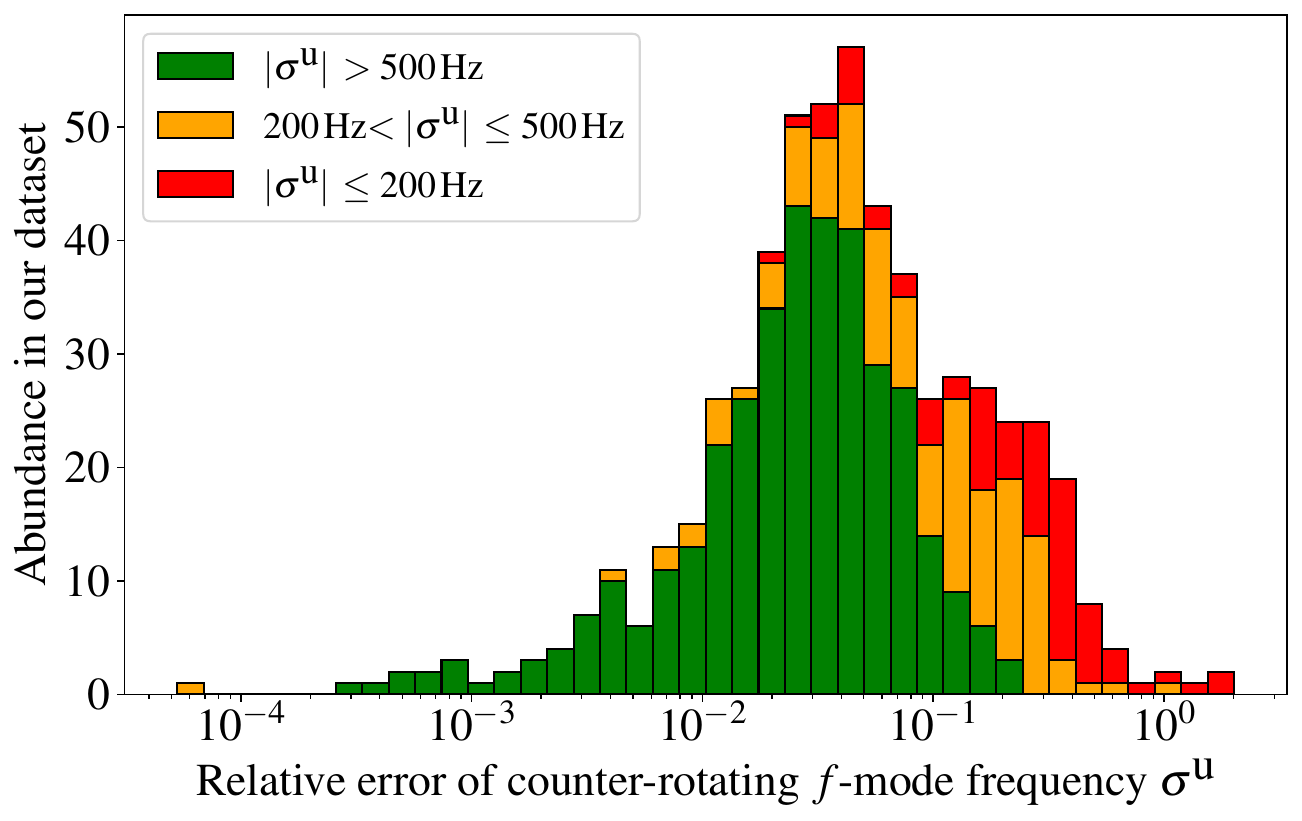}
    \caption{A stacked histogram of the relative errors as defined in Eq.~\eqref{eq:f2_rel_err} when estimating the counter-rotating $f$-mode frequency using Eq.~\eqref{eq:eq4_kk20}. Since relative errors for low frequencies are an unreliable measure for accuracy, we split the set of relative errors into three groups depending on the absolute value of the $f$-mode frequency (as extracted from the time evolution). When the magnitude of the $f$-mode is larger than $500\,$Hz (shown in green), the relative errors peak at $3\%$ and are larger than $10\%$ only in a few cases. For smaller magnitudes of the $f$-mode frequency (shown in orange and red), the relative error naturally increases. However, there are still many instances where even low-frequency $f$-modes are estimated with only a small relative error.
    \label{fig:f2p_rel_err}}
\end{figure}

We use this method for approximating the $f$-mode frequency of the potentially unstable branch; we show a histogram of the relative errors in Fig.~\ref{fig:f2p_rel_err} which we define as
\begin{align}
    \delta \sigma^\textrm{x} = \frac{\left|\sigma_\mathrm{TE}^\textrm{x} - \sigma_\mathrm{UR}^\textrm{x}\right|}{\sigma_\mathrm{TE}^\textrm{x}}.
    \label{eq:f2_rel_err}
\end{align}
Here, $\sigma_\mathrm{TE}$ and $\sigma_\mathrm{UR}$ are the frequencies as extracted from the time evolution and determined by the universal relations (using Eq.~(6) from Ref.~\cite{2010ApJ...714.1234L} to estimate $\sigma_\star$ and then Eq.~(4) from Ref.~\cite{Kruger:2019zuz}), respectively. We note that the histogram in Fig.~\ref{fig:f2p_rel_err} is built on the dataset underlying Ref.~\cite{Kruger:2019zuz} with two modifications; first, we omit those results that are based on simple polytropes, second, we have, meanwhile, extended our dataset to include $f$-mode frequencies of neutron star models based on the piecewise polytropic approximations to the EOSs DDME2~\cite{2005PhRvC..71b4312L}, FSU2~\cite{2014PhRvC..90d4305C}, MPA1~\cite{Muther:1987xaa}, and MS1~\cite{1996NuPhA.606..508M} (the PP coefficients can be found in Refs.~\cite{Read:2008iy, 2019PhRvD..99l3026K}). The updated dataset contains roughly 50\% more $f$-mode frequencies than before. Even though the universal relation Eq.~\eqref{eq:eq4_kk20} is calibrated to a smaller dataset, the new frequencies are equally well approximated; this supports the validity of the detailed procedure.

Since this branch of the $f$-mode, in general, crosses the zero frequency line, relative errors will inevitably be large and of little meaning for low $f$-mode frequencies. Therefore, we show a stacked histogram in which we split our $f$-mode data into the three groups where $|\sigma^\text{u}| > 500\unit{Hz}$ (shown in green), $200\unit{Hz} < |\sigma^\text{u}| \le 500\unit{Hz}$ (shown in orange) and $|\sigma^\text{u}| \le 200\unit{Hz}$ (shown in red). The relative errors for large $f$-mode frequencies displayed in green show that the universal relation is accurate to better than $3-4\%$ in the majority of cases; only few data points have a worse accuracy. The relative errors shown in orange and red easily creep up to $100\%$ and should not be considered as a failure of the universal relation. In contrast, the absolute error of the frequency is still fairly small and does not exceed $70\,$Hz for $90\%$ of our data points.

\subsection{\texorpdfstring{Co-rotating $f$-mode frequency}{bla}}
\label{sec:fmode_cor}

As can be seen in Fig. 2 in Ref.~\cite{Kruger:2019zuz}, the co-rotating branch of the $f$-mode (which has higher frequencies than the counter-rotating branch in the inertial frame) displays a universal behaviour, too, but the spread of the data points is considerably larger than it is for the counter-rotating branch. In order to estimate the frequencies of the co-rotating branch, we will, therefore, resort to the universal relation given in Eq.~(6) in Ref.~\cite{Kruger:2019zuz}; this relation provides an estimate for the (normalised) $f$-mode frequency $\hat{\sigma}^\textrm{x}$ of a neutron star given its effective compactness $\eta$ and (normalised) angular rotation rate $\hat{\Omega}$ by virtue of
\begin{align}
    \hat{\sigma}^\textrm{x} =  \left(c^\textrm{x}_1 + c^\textrm{x}_2 \hat{\Omega} + c^\textrm{x}_3\hat{\Omega}^2 \right) +  \left(d^\textrm{x}_1 + d^\textrm{x}_3 \hat{\Omega}^2 \right) \eta.
    \label{eq:eq6_kk20}
\end{align}
We have $\hat{\sigma}^\textrm{x} = \bar{M} \sigma^\textrm{x}/\mathrm{kHz}$ and $\hat{\Omega} = \bar{M} \Omega/\mathrm{kHz}$, where $\bar{M} = M/M_\odot$. The effective compactness $\eta$ is related to the mass $M$ and moment of inertia $I$ of the star via $\eta = \sqrt{\bar{M}^3/I_{45}}$, with $I_{45} = I/10^{45} \mathrm{g \, cm^2}$. The numerical values of the coefficients $c^\textrm{x}$ and $d^\textrm{x}$ are reported in Ref.~\cite{Kruger:2019zuz} and additional information can be found in Ref.~\cite{Kruger:2021zta}.

\begin{figure}
    \includegraphics[width=1.0\linewidth]{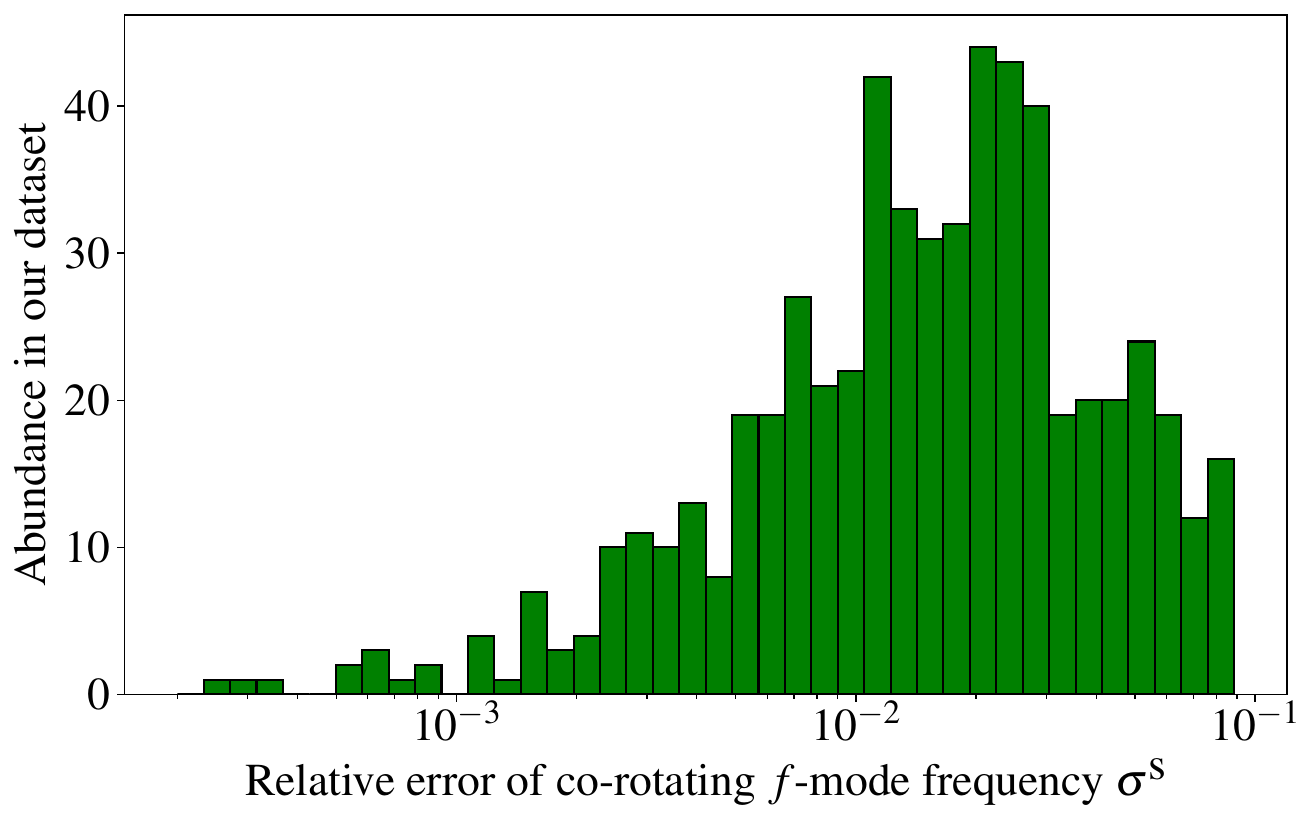}
    \caption{A histogram of the relative errors when estimating the co-rotating branch of the $f$-mode using Eq.~\eqref{eq:eq6_kk20}; we use the newly proposed universal relation for the moment of inertia and the universal relation for the mass from Ref.~\cite{2022ApJ...934..139K} to estimate the required bulk quantities of the rotating stars. We calculate the relative error as given in Eq.~\eqref{eq:f2_rel_err}. For the majority of neutron star models, our estimate is accurate to $2-3\%$; in general, the relative error stays below $10\%$.}
    \label{fig:f2m_rel_err}
\end{figure}

In order to make use of the universal relation Eq.~\eqref{eq:eq6_kk20}, we need to know the mass and moment of inertia of the rotating configuration (besides its rotation rate which in our workflow is treated as a parameter). Instead of constructing a rotating neutron star via the \textsc{rns} code, we can obtain an accurate estimate for those two quantities via the universal relation proposed in Ref.~\cite[Eq. (4)]{2022ApJ...934..139K} for the mass and using Eqs.~\eqref{eq:ur_Omega_n} - \eqref{eq:universal_x_y} for the moment of inertia. Estimating the frequency of the co-rotating $f$-mode using these universal relations results in an accuracy of better than $10\%$ but, in fact, for the majority of cases we find an accuracy of $2-3\%$; we show a histogram of the relative errors in Fig.~\ref{fig:f2m_rel_err}. Like in the case of the counter-rotating $f$-mode (cf. Sec.~\ref{sec:fmode_ctr}), we also determined a similar number of co-rotating $f$-mode frequencies based on the same EOSs. These data points extend the previous dataset on which Ref.~\cite{Kruger:2019zuz} was built; again, the universal relation, which was fitted to the smaller dataset, estimates the newly determined $f$-mode frequencies well.

In this section, we have compared $f$-mode frequencies as obtained from time evolutions to estimates from simple analytical formulae. As can be seen in Figs.~\ref{fig:f2p_rel_err} and \ref{fig:f2m_rel_err}, the relative error is at percent level and in only a few cases exceeds $10\%$. Note that we achieved this accuracy for the $f$-mode frequencies of a star spinning with angular frequency $f_\textrm{spin}$ out of the three numbers $M_\star$, $R_\star$, and $I_\star$.

\subsection{The EOS inference code}

In a recent work~\cite{Volkel:2022utc}, we presented a Bayesian framework that uses neutron star observables to infer the Read \textit{et al.}~\cite{Read:2008iy} piecewise polytropic EOS from a precomputed TOV template bank. 
In that framework, rotational effects have only partially been included, i.e., by using the co- and counter-rotating $f-$mode frequencies, but neglecting rotational effects for other neutron star properties that enter at quadratic order. 
However, as input data, rotational effects had been taken into account properly in order to represent a more realistic scenario and quantify possible bias. 
For more details on the implementation we refer to the original work~\cite{Volkel:2022utc}. 

In the present work, we build on the previously developed inference code and extend it to properly account for rotational corrections to the mass, radius and moment of inertia using the universal relations proposed in Ref.~\cite{2022ApJ...934..139K} and those presented in this paper. This will lift the former limitation of being restricted to observations of neutron stars for which oblateness can be neglected. For simplicity, we will use the same set of potentially measured observables as in the previous work, i.\,e., the mass, (circumferential, equatorial) radius, rotation rate and two $f$-mode frequencies of a rotating neutron star. We note that it does not pose any difficulty to extend the code to also allow other observables to be taken into account; in the present work, however, our focus lies on the extension to rapid rotation and to study how well EOS inference works when relying on universal relations.

\subsubsection{The EOS database}
\label{sec:eos_db}

We use essentially the same EOS database as in the preceding study; we will repeat the relevant details here but refer the reader to Sec.~D in Ref.~\cite{Volkel:2022utc} for additional details.

During sampling, MCMC walkers will move through the four-dimensional EOS parameter space which is spanned by the PP parameters $\theta = (p_1, \Gamma_1, \Gamma_2, \Gamma_3)$, for which we assume flat priors. We confine these four parameters by intervals such that all best fits provided in Table~III in Ref.~\cite{Read:2008iy} are covered. The intervals are given by
\begin{align}
    \log_{10}(p_1) & \in [33.940, 34.860 ], \\
    \Gamma_1  & \in [ 2.000, 4.070 ], \\
    \Gamma_2  & \in [ 1.260, 3.800 ], \\
    \Gamma_3  & \in [ 1.290, 3.660 ],
\end{align}
where $p_1$ is given in $\text{dyn}/\text{cm}^2$. An MCMC walker needs to calculate the likelihood of making a step to a new point $\theta'$ in the parameter space. Ideally, we would calculate the $M$-$R$ curve of the EOS specified by $\theta'$ by solving the TOV equations on the fly, however, this is despite the simplicity of the TOV equations computationally somewhat expensive. Therefore, we build a large EOS database for which we divide each of the above intervals for the four PP parameters onto an evenly spaced grid of $N_\textrm{EOS}$ points and calculate the $M$-$R$ curves (including the moment of inertia $I$) for each grid point. The likelihood of an MCMC walker making a step to $\theta'$ will be calculated with respect to the EOS $\theta_\textrm{approx}$, where $\theta_\textrm{approx}$ is the EOS present in the database that is closest to $\theta'$ (in the sense of being the nearest neighbour). In order to perform convergence tests, we build databases for $N_\textrm{EOS} = 30, 40, 50$. Note that all our results are based on the $N_\textrm{EOS}=50$ resolution.

\subsection{Bayesian inference on EOS}
\label{sec:bayes_inf}

In the following, we compare results of this framework before and after including rotational effects using the previously derived universal relations. 
This upgrade promotes our framework to model arbitrary rotation rates for neutron star bulk properties from the same non-rotating TOV template bank as before. 

For the sake of clarity, we focus on a limited set of examples to demonstrate when the improvements become crucial. 
In all subsequent examples we have used the \textsc{python} MCMC sampler \textsc{emcee}~\cite{Foreman-Mackey:2012any} with 100 walkers, 2000 steps and in total 12 parallel runs. 
The sampling of each of the following examples only takes about $10-20\,$min on a standard laptop. 
Posterior corner plots have been constructed using the \textsc{python} package \textsc{corner}~\cite{corner}. 
We strongly emphasize that all mock data used in the following, as done in Ref.~\cite{Volkel:2022utc}, have been produced with the \textsc{rns} code and the $f$-mode frequencies have been directly taken from the time evolution presented in Ref.~\cite{Kruger:2019zuz}. 
We consider the SLy EOS~\cite{Douchin:2001sv} and noiseless injections of neutron star observables with a relative error of $5\%$, as our main objective here is to study the performance of Bayesian framework, and not to carry out astrophysical projections of future measurements. 

\begin{figure*}
\includegraphics[width=0.495\linewidth]{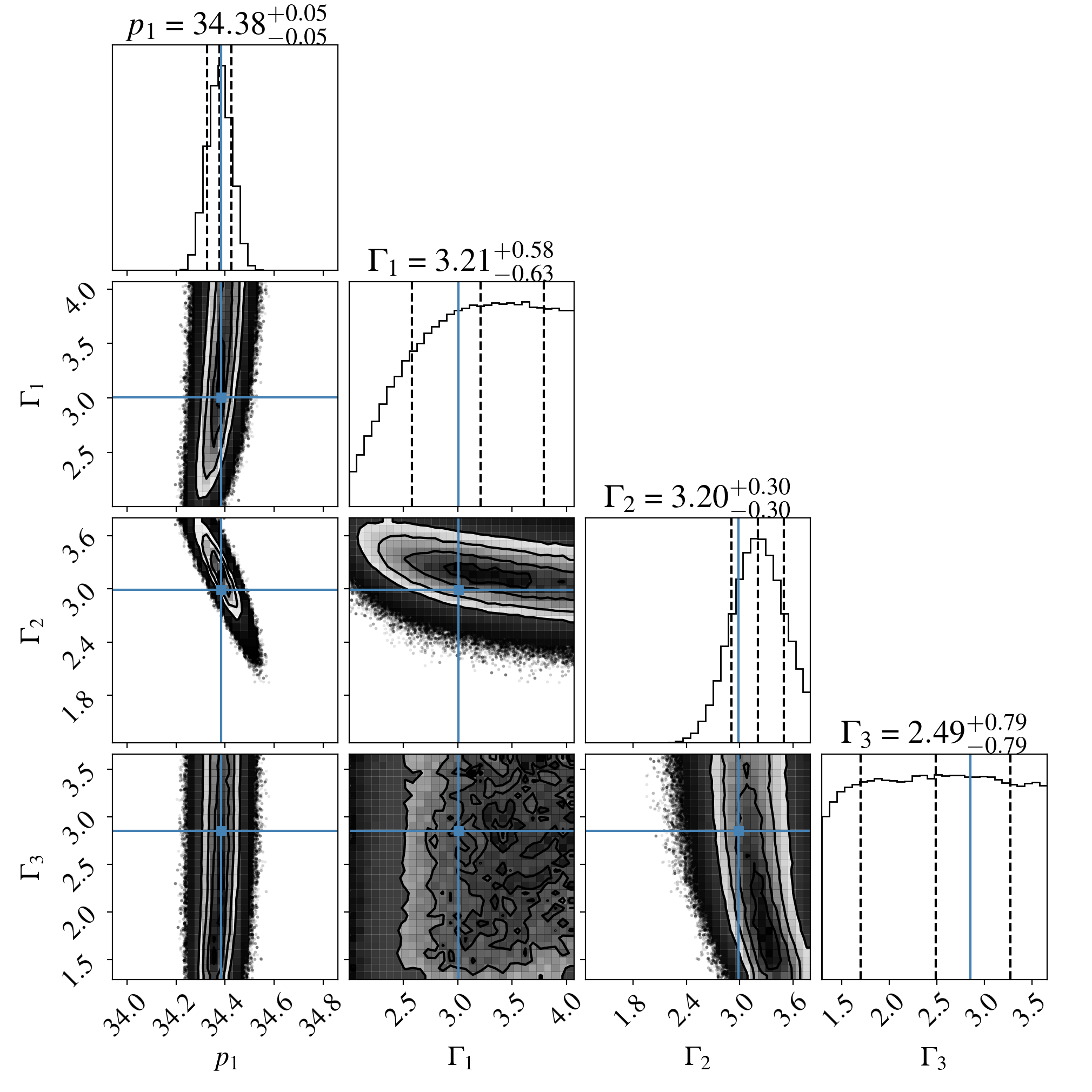}
\includegraphics[width=0.495\linewidth]{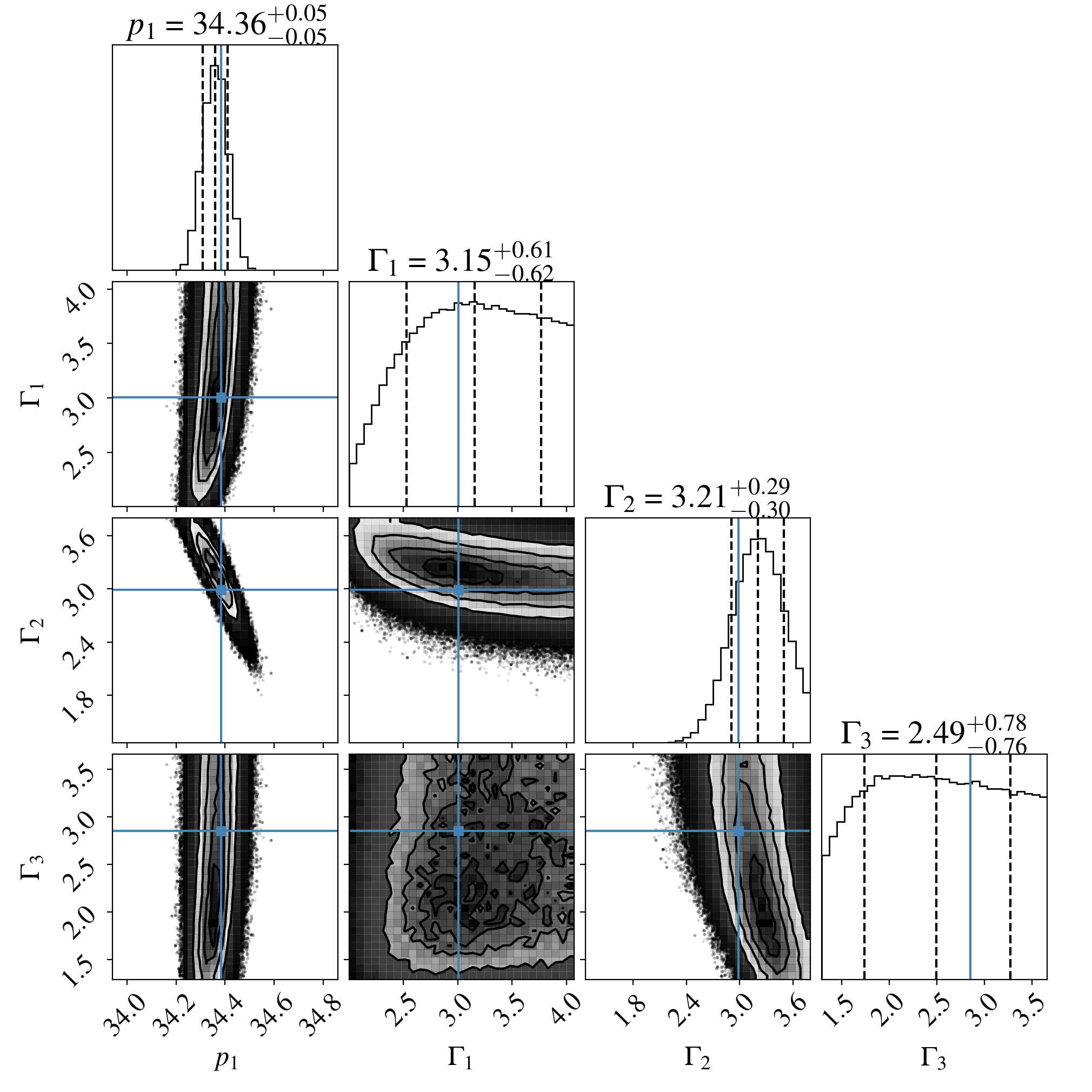}
\caption{Results of the MCMC analysis using observations of slowly rotating neutron stars listed in Table~\ref{tab:coeff_SLy_slowly} as described in the main text. The two corner plots compare the results using the old framework with slow-rotation approximations (left panel) and the new framework based on universal relations that allow rapid rotation (right panel).
The blue lines indicate the injected EOS parameters and black dashed intervals define the \{0.16, 0.5, 0.84\} quantils. The results of both frameworks are almost identical which reconfirms that the slow-rotation approximation is justified for observations of neutron stars that rotate sufficiently slowly (cf.~Ref.~\cite{Volkel:2022utc}).}
\label{mcmc_1}
\end{figure*}

First, we study a set of slowly rotating neutron stars for which the mock data can be found in Table~\ref{tab:coeff_SLy_slowly}. 
In Fig.~\ref{mcmc_1} we show that the EOS posteriors as obtained in our previous work (left panel), which also assumed a set of slowly rotating neutron stars, are almost identical to the ones obtained with the new universal relations (right panel). 
This further quantifies that the slow-rotation approximations are justified for such cases. 
Because none of the neutron stars is particularly heavy, the high-density regime of the EOS is not explored well, which agrees with the rather uninformed posteriors for $\Gamma_3$. 

\begin{table}
    \caption{Mock data for slowly rotating neutron stars constructed with the SLy EOS.}
    \label{tab:coeff_SLy_slowly}
    \begin{ruledtabular}
    \begin{tabular}[v]{ccccccc}
        n & $M$ & $R_e$ & $f_{\rm spin}$ & $\sigma^{\rm u}$ & $\sigma^{\rm s}$ & $\mathfrak{r}$\\
          & $[M_\odot]$ & $[\mathrm{km}]$ & $[\mathrm{kHz}]$ & $[\mathrm{kHz}]$ & $[\mathrm{kHz}]$ &  \\
        \hline
        1    & 1.282	& 11.827 & 0.192 & 1.590 & 2.075 & 0.990 \\
        2    & 1.200	& 11.902 & 0.238 & 1.486 & 2.082 & 0.983 \\
        3    & 1.363	& 11.792 & 0.211 & 1.603 & 2.143 & 0.989 \\
        4    & 1.924	& 11.013 & 0.272 & 1.862 & 2.649 & 0.990 \\
        5    & 1.601	& 11.637 & 0.283 & 1.618 & 2.368 & 0.984 \\
        6    & 1.788	& 11.340 & 0.247 & 1.798 & 2.476 & 0.990 \\
    \end{tabular}
    \end{ruledtabular}
\end{table}

\begin{figure*}
\includegraphics[width=0.495\linewidth]{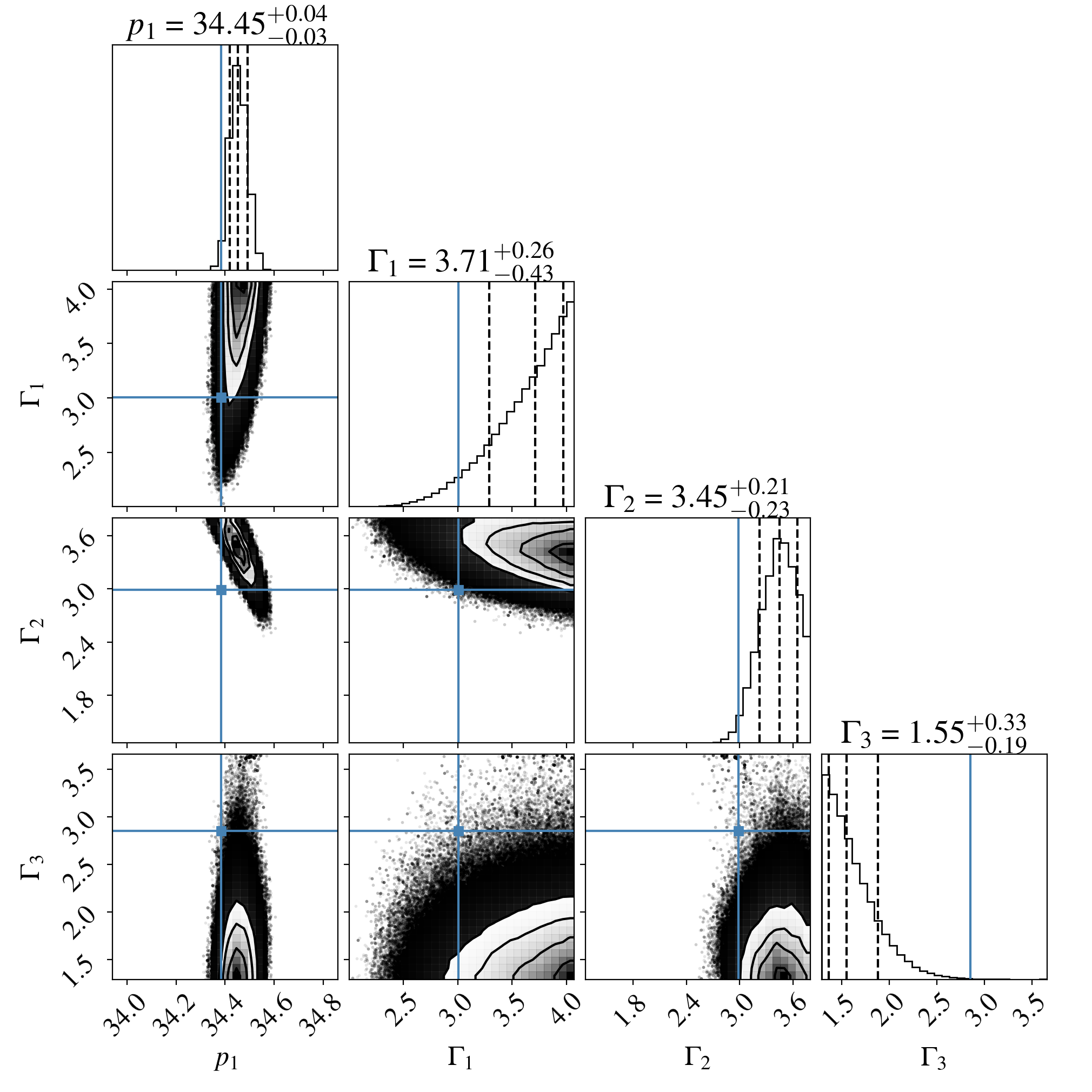}
\includegraphics[width=0.495\linewidth]{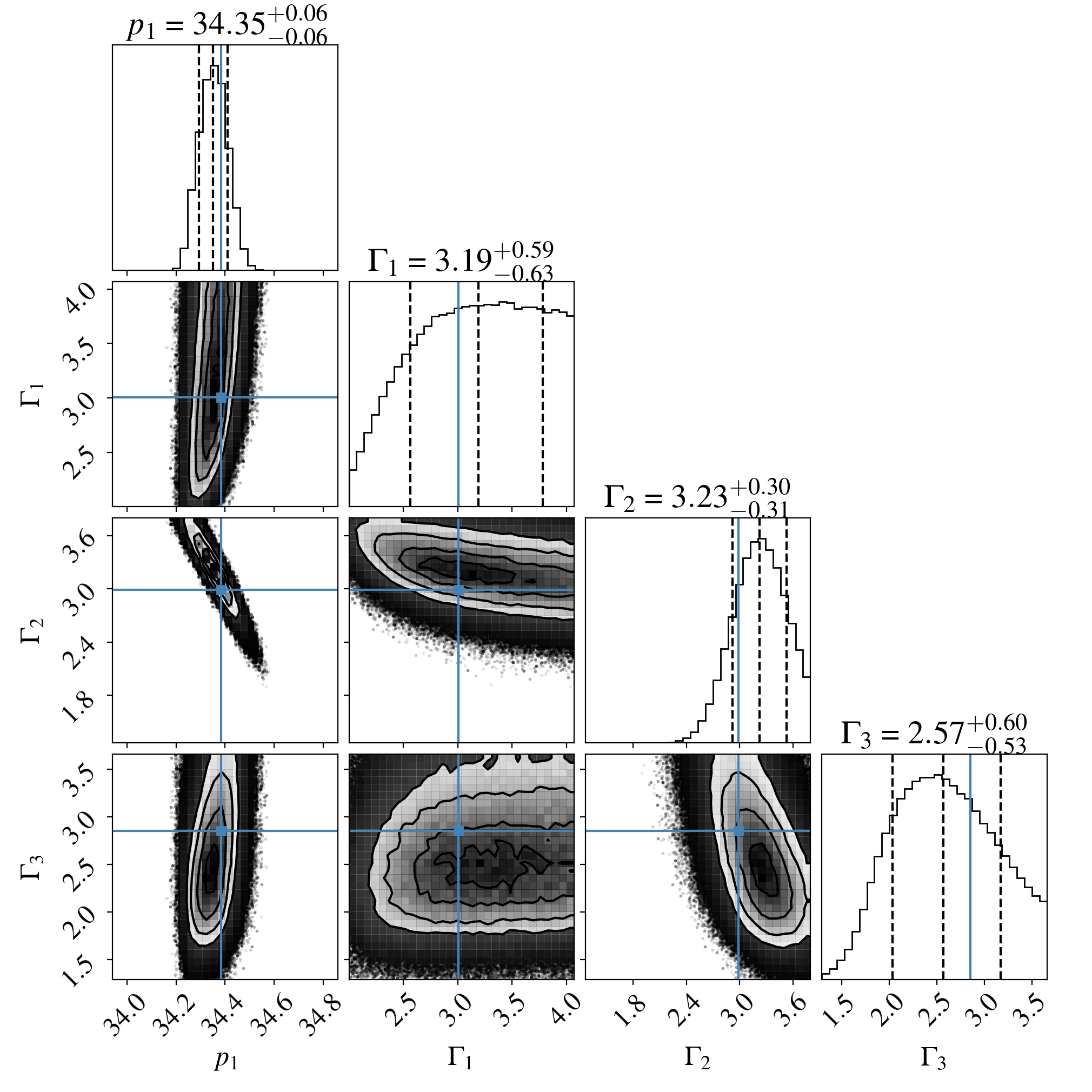}
\caption{The same as Fig.~\ref{mcmc_1} but for observations of moderately rotating neutron stars as shown in Table~\ref{tab:coeff_SLy_moderately}. At rotation rates  of $350 - 550\,$Hz or axis ratios of $\mathfrak{r} \approx 0.96 - 0.97$, the stars' oblateness is no longer negligible and we observe a strong bias in the posteriors, in particular for $\Gamma_1$ and $\Gamma_3$, when using the old framework. The new framework reconstructs the PP parameters well.}
\label{mcmc_2}
\end{figure*}

Next, we consider a set of moderately rotating neutron stars for which the mock data can be found in Table~\ref{tab:coeff_SLy_moderately}. 
The corresponding EOS posteriors from the MCMC analysis are reported in Fig.~\ref{mcmc_2}. 
It is evident that the EOS parameters obtained with the slow-rotation approximation are strongly biased, while the ones obtained with the universal relation are comparable to the previous case in Fig.~\ref{mcmc_1}. 
Because some of the considered neutron stars in this dataset are heavier compared to the previous one, the high-density constraints on $\Gamma_3$ are much more informative, while the other EOS parameters are constrained similarly well. 
For similar findings see also Ref.~\cite{Weih:2019rzo}. 

\begin{table}
    \caption{Mock data for moderately rotating neutron stars constructed with the SLy EOS.}
    \label{tab:coeff_SLy_moderately}
    \begin{ruledtabular}
    \begin{tabular}[v]{ccccccc}
        n & $M$ & $R_e$ & $f_{\rm spin}$ & $\sigma^{\rm u}$ & $\sigma^{\rm s}$ & $\mathfrak{r}$\\
          & $[M_\odot]$ & $[\mathrm{km}]$ & $[\mathrm{kHz}]$ & $[\mathrm{kHz}]$ & $[\mathrm{kHz}]$ &  \\
        \hline
       1  & 1.800 & 11.491 & 0.472 & 1.416 & 2.728 & 0.963 \\
       2  & 1.200 & 12.035 & 0.358 & 1.300 & 2.187 & 0.962 \\
       3  & 1.828 & 11.440 & 0.484 & 1.420 & 2.776 & 0.962 \\
       4  & 1.739 & 11.548 & 0.412 & 1.484 & 2.610 & 0.970 \\
       5  & 2.030 & 10.692 & 0.522 & 1.556 & 3.118 & 0.970 \\
       6  & 1.500 & 11.858 & 0.410 & 1.365 & 2.432 & 0.962 \\
    \end{tabular}
    \end{ruledtabular}
\end{table}

\begin{figure*}
\includegraphics[width=0.495\linewidth]{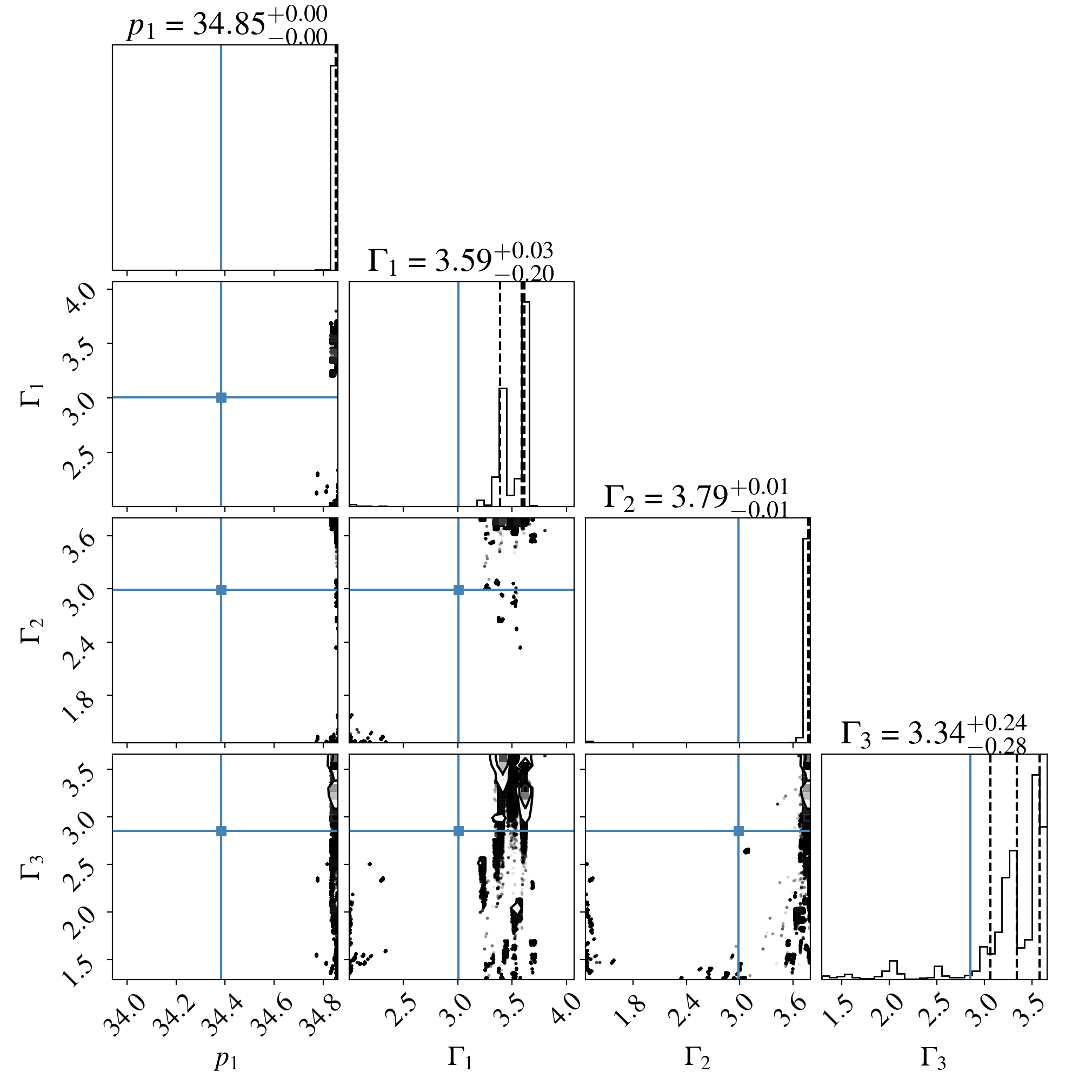}
\includegraphics[width=0.495\linewidth]{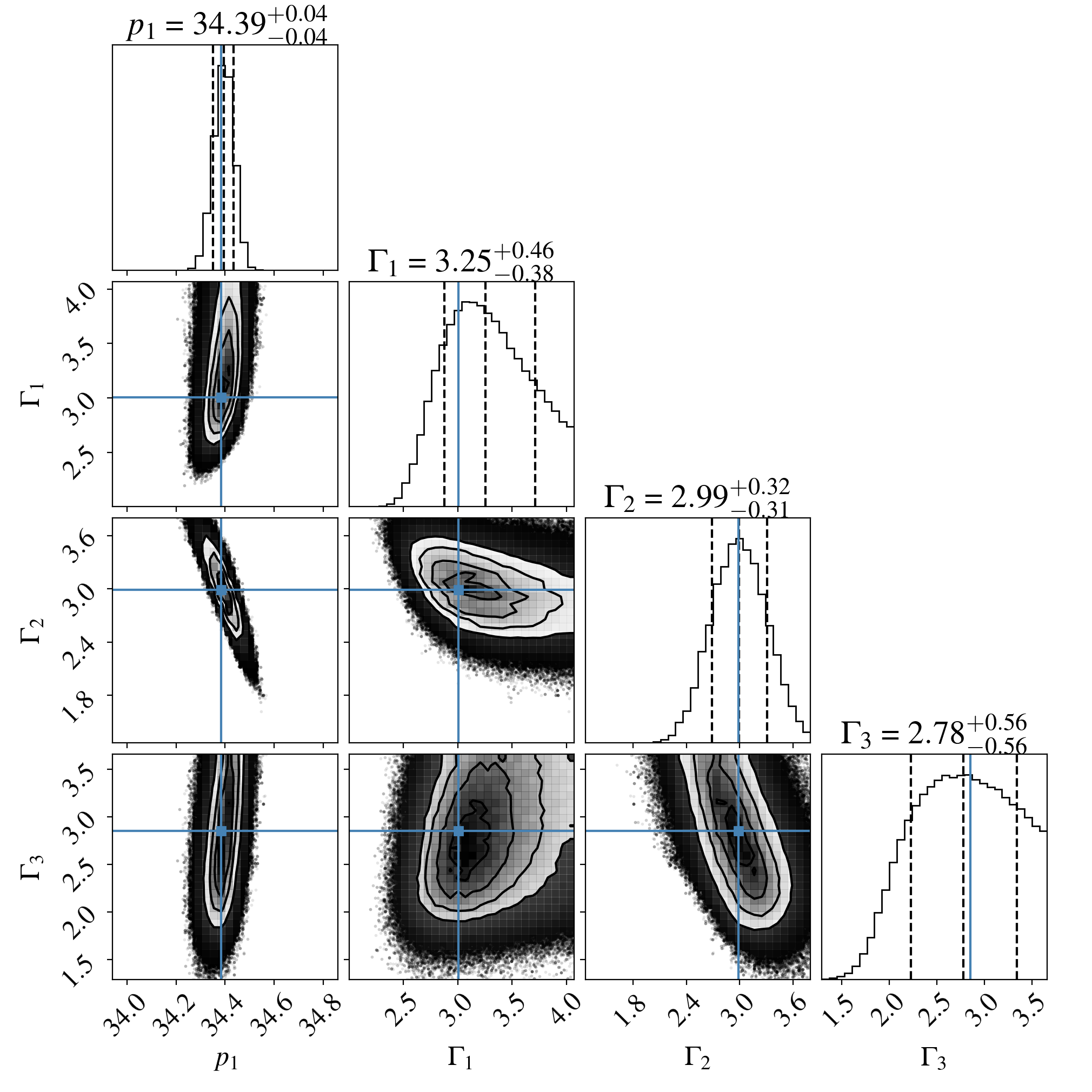}
\caption{The same as Fig.~\ref{mcmc_1} but for observations of rapidly rotating neutron stars as shown in Table~\ref{tab:coeff_SLy_rapidly}. With axis ratios of $\mathfrak{r} \approx 0.73 - 0.90$, the stars are far from sphericity and as a result, the old framework fails completely. The new framework built on the universal relations for rapid rotation, again, yields very useful results.}
\label{mcmc_3}
\end{figure*}

As final test, we consider the set of rapidly rotating neutron stars that can be found in Table~\ref{tab:coeff_SLy_rapidly}. 
The results of the MCMC parameter estimation are shown in Fig.~\ref{mcmc_3}, which is organized as the previous ones. 
It is evident that the slow-rotation approach fails completely, as the posteriors of most EOS parameters ramp up against the boundary of the allowed parameter space. 
In strong contrast are the results using the universal relations, which even in this rather extreme case yield very reliable bounds. 

\begin{table}
    \caption{Mock data for rapidly rotating neutron stars constructed with the SLy EOS.}
    \label{tab:coeff_SLy_rapidly}
    \begin{ruledtabular}
    \begin{tabular}[v]{ccccccc}
        n & $M$ & $R_e$ & $f_{\rm spin}$ & $\sigma^{\rm u}$ & $\sigma^{\rm s}$ & $\mathfrak{r}$\\
          & $[M_\odot]$ & $[\mathrm{km}]$ & $[\mathrm{kHz}]$ & $[\mathrm{kHz}]$ & $[\mathrm{kHz}]$ &  \\
        \hline
       1  & 1.200 & 13.191 & 0.744 & 0.567 & 2.419 & 0.812 \\
       2  & 2.100 & 10.899 & 0.966 & 0.747 & 3.665 & 0.900 \\
       3  & 1.379 & 13.870 & 0.912 & 0.254 & 2.582 & 0.734 \\
       4  & 1.774 & 12.678 & 0.965 & 0.380 & 3.013 & 0.809 \\
       5  & 1.467 & 13.613 & 0.927 & 0.273 & 2.665 & 0.750 \\
       6  & 1.573 & 13.005 & 0.896 & 0.413 & 2.775 & 0.800 \\
    \end{tabular}
    \end{ruledtabular}
\end{table}

Note that in all of the here presented cases, despite arguably the last one for the universal relation approach, the injected EOS parameters cannot be reconstructed perfectly. 
This is the case even though we have not added noise to the data (noiseless injections). 
Instead, the small bias is due to the remaining imperfections in adopting universal relations and cannot easily be improved in the full rotating case, but recently something related has been studied for universal relations using the tidal deformability~\cite{Kashyap:2022wzr}. 
However, as the assumed measurement errors are already very small (relative errors of $5\%$), the bias is negligible but may become relevant for more precise measurements. 
Although all the here presented results have been obtained for the SLy EOS, we note that we have also performed EOS inference runs using observations based on the MPA1 EOS~\cite{Muther:1987xaa}. 
We find qualitatively very similar behaviour for the robustness of the universal relation approach, and clear failure of the slow-rotation approach when applied to rapidly rotating neutron stars. 
More specifically, the true values of all EOS parameters are well within the \{0.16, 0.84\} quantils, and the constraints on $p_1$ and $\Gamma_2$ are more informative than the ones for $\Gamma_1$ and $\Gamma_3$. Note that the MPA1 EOS was not used in Ref.~\cite{Kruger:2019zuz} for the construction of the universal $f$-mode relation.

\begin{figure}
\includegraphics[width=1.0\linewidth]{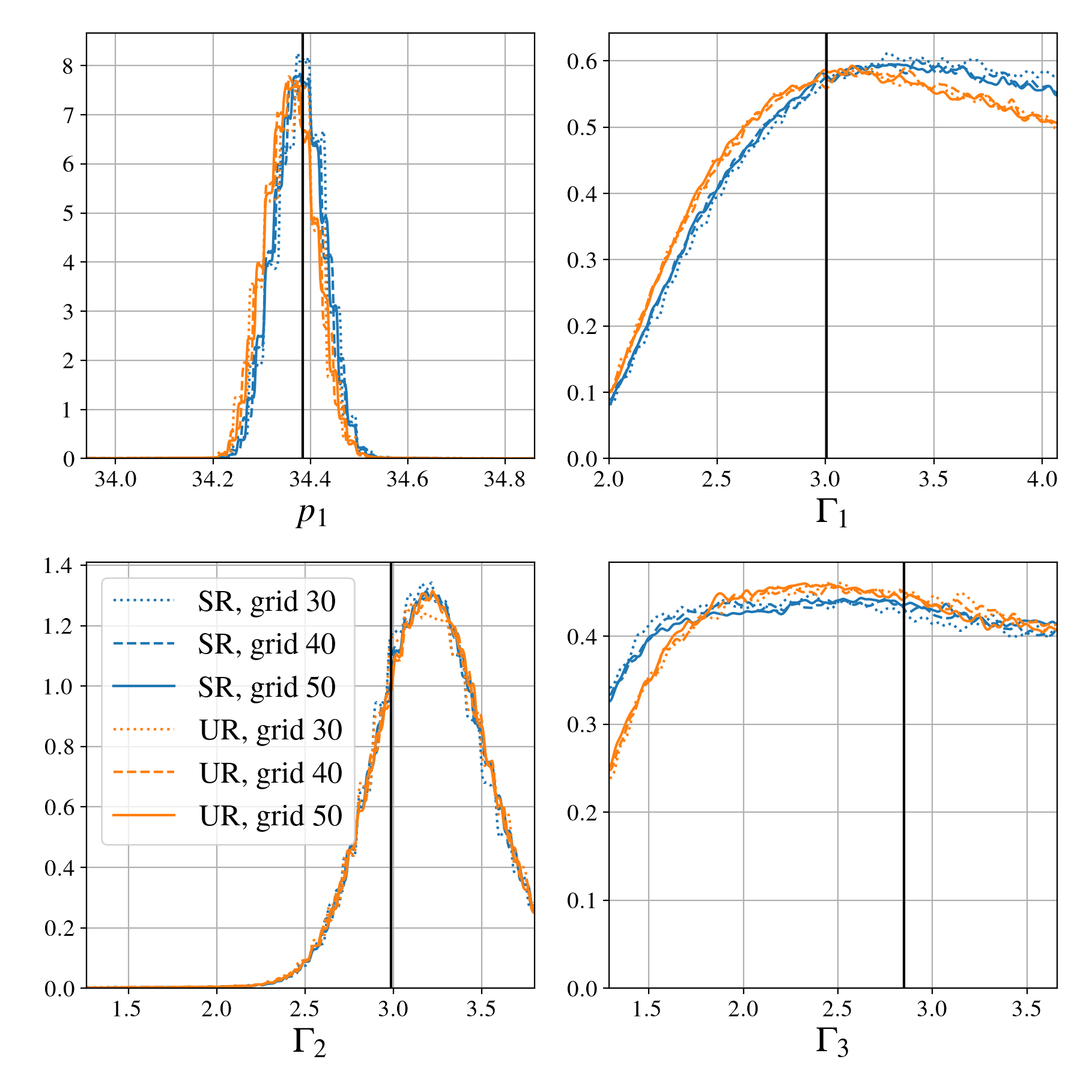}
\caption{Here we show MCMC results using different EOS database resolutions for slowly rotating neutron star observables from Table~\ref{tab:coeff_SLy_slowly}. 
The slow-rotation approximation (SR, blue) and universal relation (UR, orange) methods have been applied with EOS grid resolutions of {30, 40, 50} (different line styles). 
Black vertical lines indicate the correct EOS parameters.
\label{eos_res}
}
\end{figure}

\begin{figure}
\includegraphics[width=1.0\linewidth]{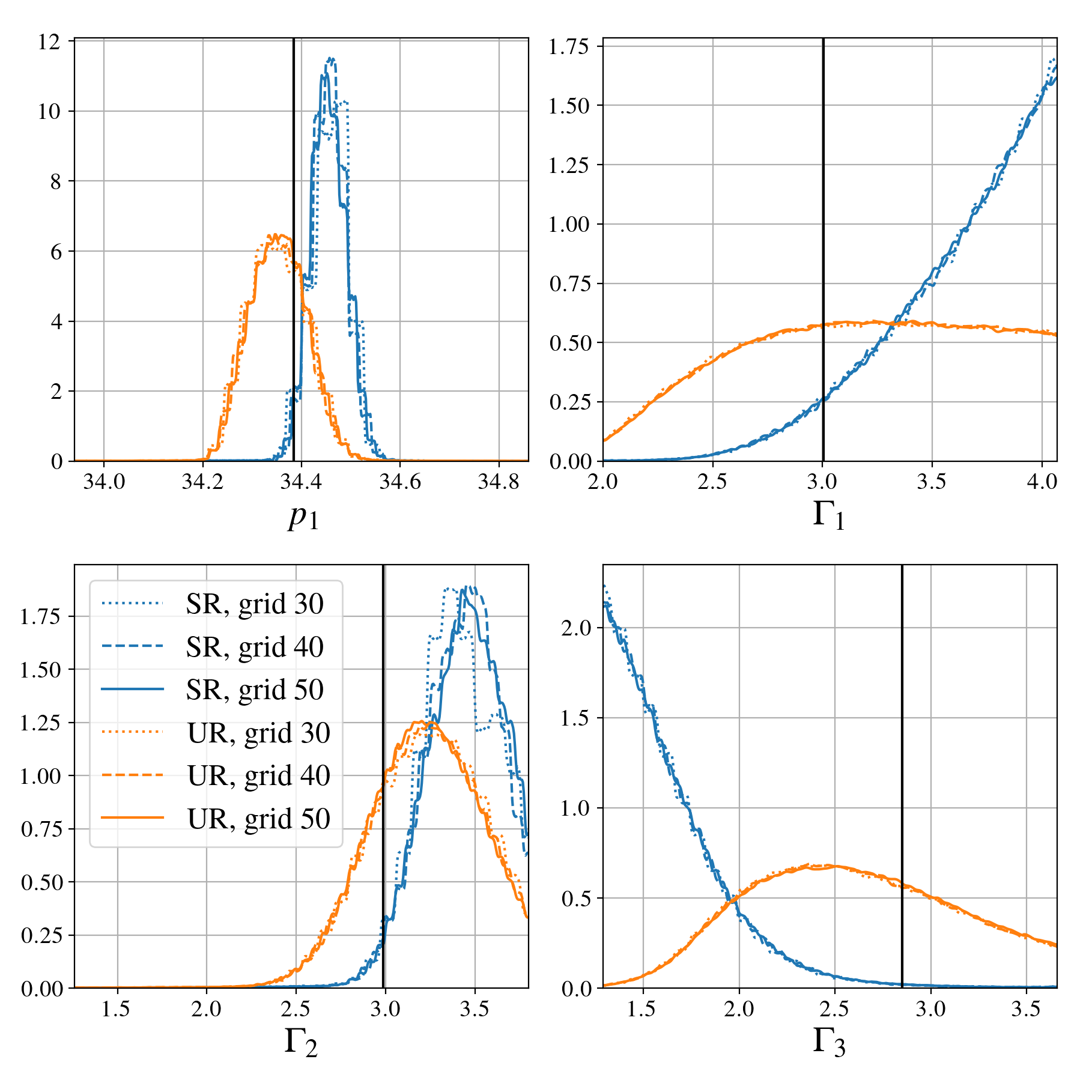}
\caption{Here we show MCMC results using different EOS database resolutions for the moderately rotating ones from Table~\ref{tab:coeff_SLy_moderately}. 
The chosen resolutions and definition of lines are the same as those in Fig.~\ref{eos_res}. \label{eos_res2}}
\end{figure}

To verify to what extend our results depend on the resolution of the precomputed EOS database, we have conducted several convergence tests. 
One of them is provided in Fig.~\ref{eos_res} and shows the 1D posteriors distributions of the EOS parameters when sampled with different database resolutions, which are defined by 30, 40, and 50 steps for each parameter. 
Here we show results for slowly rotating neutron star observables from Table~\ref{tab:coeff_SLy_slowly}. 
In Fig.~\ref{eos_res2}, the ones for moderately rotating data from Table~\ref{tab:coeff_SLy_moderately} can be found. 
Both cases demonstrate that the posteriors are very robust with respect to the different resolutions, but small artefacts of the EOS grid resolution can be seen, e.g., the steplike structure for $p_1$ for the lowest resolution. 
In Fig.~\ref{eos_res}, the direct comparison between the slow-rotation approximation and universal relation based results indicate a small, but consistent bias for $p_1$, as well as traces of one for $\Gamma_1$, which are both more relevant for the low-density region of the neutron star. 
$\Gamma_2$ is very similar, indicating that it is less sensitive to rotational effects, which should be expected. 
Since $\Gamma_3$ is not well probed by most of the data, the posteriors are flat and the possible presence of bias is less relevant. 
In Fig.~\ref{eos_res2}, one clearly sees that the chosen resolution is much less important compared to the significant biases that are introduced through the slow-rotation approximation. 
Furthermore, we have also studied even lower resolutions of {10, 20} (not shown), but conclude that they are not fine enough to provide robust answers for the here studied relative errors of $5\%$. 
Finally, we also studied the \{0.16, 0.5, 0.84\} quantils for different resolutions and find, as expected from Figs.~\ref{eos_res} and~\ref{eos_res2}, that these quantities are very robust, as the coarse grid resolution tends to average out. 

The main conclusions from these demonstrations are clear. 
Modeling significantly rotating neutron star observables with a slow-rotation approximation framework is, as expected, in general not robust and can cause major bias in the reconstructed EOS parameters when violated. 
However, using the new universal relations yields robust results, even when the provided observables have been provided with a few percent-level uncertainty. 
We note that due to the simple analytic structure of the universal relations, there is no significant increase in computational time during sampling (depending on resolution and other details around a factor of 2). 
This would be drastically different, if the stellar structure equations would have to be solved during sampling.

\section{Conclusions}
\label{sec:conclusions}

We have presented four universal relations that allow to approximate some of the bulk properties of uniformly rotating neutron stars at percent-level accuracy. 
In particular, we propose universal relations to obtain the moment of inertia $I$ and the value of $T/W$ at the Kepler limit and subsequently also for arbitrarily fast rotating neutron stars.
These universal relations are insensitive to the EOS and rely solely on the knowledge of the triple $(M_\star, R_\star, I_\star)$ of the zero-spin star with the same central energy density, which can easily be calculated by solving the TOV equations along with Hartle's equation. We also explain how the estimated quantities $I$ and $T/W$ can be combined with other known or estimated quantities in order to obtain estimates for the angular momentum $J$, the rotational kinetic energy $T$, the gravitational binding energy $W$ and the proper mass $M_p$.

We built a large dataset of 1550 sequences of constant central energy density across 31 EOSs. The universal relations for $I$ and $T/W$ are calibrated using this dataset. The data are strongly correlated and we find that the quantities as computed using the polynomial fits deviate on average less than 0.6\% from the \textsc{rns} code; the maximum relative errors of the four universal relation lies between $1.8\%$ and $4.3\%$.
Combining our results with recent universal relations for the mass and equatorial radius from Konstantinou and Morsink~\cite{2022ApJ...934..139K}, we have considered two applications to demonstrate the potential for future applications. 

First, using the universal relations for mass and moment of inertia, we have computed approximations to the frequencies of the co- and counter-rotating $l=|m|=2$ $f$-modes from non-rotating neutron star properties only and compared them to values obtained from time evolutions in an earlier work (cf.~Refs.~\cite{Kruger:2019zuz, Kruger:2020ykw}). We find very good agreement and the frequency estimates yield an accuracy of $2-3\%$ in the majority of cases; only a few $f$-mode frequencies deviate more than $10\%$ from our estimate.

Second, we have incorporated the new universal relations into the Bayesian framework presented in Ref.~\cite{Volkel:2022utc}. 
This allows us to infer the parameters of the Read \textit{et al.}~\cite{Read:2008iy} EOS model from a hypothetical set of neutron star observations beyond slow-rotation approximations for the statistical modeling. 
For our inference analysis, we use three different sets of mock observations which we take from the dataset of our earlier work \cite{Kruger:2019zuz}: one set of slowly rotating neutron stars such that their oblateness is negligible, one set of moderately rotating stars for which their axis ratios are $\mathfrak{r} \approx 0.96 - 0.97$, and one set of fast rotating stars with axis ratios of $\mathfrak{r} \approx 0.73 - 0.90$. 
In all presented cases, the use of universal relation-based modeling is sufficient to reconstruct the injected EOS parameters within \{0.16, 0.84\} quantils. 
Our inference analysis also demonstrates that using slow-rotation approximations for sets of moderately and fast rotating neutron star observables yields significant bias, and that using the new universal relations is crucial.
Our choice of mock observations results in more informative posteriors for the EOS piecewise polytropic indices at high densities, when considering a similar number of observables with similar measurement accuracy. 
Because of the very fast numerical evaluation of the underlying universal relations, modeling rotating stars has negligible impact on computational time compared to the slow-rotation scheme used earlier~\cite{Volkel:2022utc}. 
We emphasize that the possibility to avoid solving the Einstein equations for rotating neutron stars during sampling or to avoid building a significantly larger precomputed EOS database provides major improvements for future parameter estimation studies. 

In the EOS inference runs, we use mass, equatorial radius, rotation rate and two $f$-mode frequencies as observables. The new universal relation for the moment of inertia is essential in the estimation of the latter; as our focus lies on the extension of a previous code to rapid rotation, we do not consider $I$ or $T/W$ to be observables in our inference runs. However, it poses no difficulty to incorporate these quantities observables into the inference code employing the respective universal relations, should the need arise.

Future work may also include the extension to more realistic models for the underlying EOS; this means employing different kinds of EOS parametrisations to broaden the covered range of EOSs (and reduce a potential bias) but also to take into account (astro)physical constraints to narrow the range of EOSs. Furthermore, the possibility of phase transitions opens up a different pathway for potential extensions.
\newline

\acknowledgments
The authors thank Tim Dietrich, Erich Gaertig, Kostas Glampedakis, Kostas D. Kokkotas, André Oliva and Luciano Rezzolla for useful discussions. The authors also thank the anonymous referees for useful comments which helped improve the manuscript.
S.~H.~V\"olkel acknowledges funding from the Deutsche Forschungsgemeinschaft (DFG) - project number: 386119226.

\end{document}